\documentclass[fleqn,usenatbib]{mnras}

\usepackage{newtxtext,newtxmath}

\usepackage[T1]{fontenc}
\usepackage{ulem}

\DeclareRobustCommand{\VAN}[3]{#2}
\let\VANthebibliography\thebibliography
\def\thebibliography{\DeclareRobustCommand{\VAN}[3]{##3}\VANthebibliography}

\usepackage{graphicx}	
\usepackage{amsmath}	

\title[The evolution of satellite galaxies]{Tracking the evolution of satellite galaxies: mass stripping and dark-matter deficient galaxies}

\author[Montero-Dorta et al.]{
\parbox[t]{\textwidth}{
Antonio D. Montero-Dorta$^{1}$\thanks{E-mail: antonio.montero@usm.cl}, Facundo Rodriguez$^{2,3}$, M. Celeste Artale$^{4,5,6}$, Rory Smith$^{1}$, Jon\'as  Chaves-Montero$^{7,8}$} 
\vspace*{6pt} \\ 
$^{1}$  Departamento de F\'isica, Universidad T\'ecnica Federico Santa Mar\'ia, Casilla 110-V, Avda. Espa\~na 1680, Valpara\'iso, Chile\\
$^{2}$ CONICET. Instituto de Astronom\'ia Te\'orica y Experimental. Laprida 854, X5000BGR, C\'ordoba, Argentina \\
$^{3}$ Universidad Nacional de C\'ordoba. Observatorio Astron\'omico de C\'ordoba. C\'ordoba, Argentina \\
$^{4}$ Physics and Astronomy Department Galileo Galilei, University of Padova, Vicolo dell'Osservatorio 3, I--35122, Padova, Italy\\
$^{5}$ INFN--Padova, Via Marzolo 8, I--35131 Padova, Italy\\
$^{6}$ Department of Physics and Astronomy, Purdue University, 525 Northwestern Avenue, West Lafayette, IN 47907, USA\\
$^7$ Donostia International Physics Center (DIPC), Paseo Manuel de Lardiz\'abal, 4, 20018 Donostia-San Sebasti\'an, Spain \\
$^8$ Institut de F\'isica d'Altes Energies, The Barcelona Institute of Science and Technology, Campus UAB, E-08193 Bellaterra (Barcelona), Spain \\
\vspace{-0.4cm} 
}

\date{Accepted XXX. Received YYY; in original form ZZZ}

\pubyear{2022}

\begin{document}
\label{firstpage}
\pagerange{\pageref{firstpage}--\pageref{lastpage}}
\maketitle

\begin{abstract}
 
Satellite galaxies undergo a variety of physical processes when they are accreted by groups and clusters, often resulting in the loss of baryonic and dark matter (DM) mass. In this work, we evaluate the predictions from the IllustrisTNG hydrodynamical simulation regarding the evolution of the matter content of satellites, focusing on a population that are accreted at $z>1$ and retain their identity as satellites down to $z=0$. At fixed host halo mass, the amount of DM and stellar mass stripped depends mostly on the pericentric distance, $d_{\rm peri}$, here normalised by host halo virial radius. The closest encounters result in significant loss of DM, with subhaloes retaining between 20 and a few per cent of their $z=1$ mass. At fixed $d_{\rm peri}$, DM mass stripping seems more severe in lower mass haloes. Conversely, the average satellite in higher mass haloes has its stellar mass growth halted earlier, having lost a higher fraction of stellar mass by $z=0$. We also show that mass stripping has a strong impact on the quenched fractions. The IllustrisTNG boxes are qualitatively consistent in these predictions, with quantitative differences mostly originating from the distinct subhalo mass ranges covered by the boxes. Finally, we have identified DM-deficient systems in all TNG boxes. These objects are preferentially found in massive clusters ($M_{\rm host } \gtrsim 10^{13}$ M$_\odot$), had very close encounters with their central galaxies ($d_{\rm peri}\simeq0.05\, R_{\rm vir}$), and were accreted at high redshift ($z_{\rm infall} \gtrsim 1.4$), reinforcing the notion that tidal stripping is responsible for their remarkable lack of DM.

\end{abstract}




\section{Introduction}\label{sec:Intro}

In the standard $\Lambda$CDM cosmological model, galaxies form by the cooling and condensation of baryons at the bottom of the potential wells of cold dark matter (DM) haloes \citep{WhiteRees1978, WhiteFrenk1991}. This galaxy formation process is highly nonlinear and involves a great variety of astrophysical mechanisms over a wide range of scales. Despite this complexity and the distinct nature of the DM and the baryonic components, several connections have been established between the properties of galaxies and the properties of their hosting haloes. As an example, central galaxies in massive haloes are typically large and have high stellar contents, as dictated by the galaxy size — halo mass relation (e.g., \citealt{Kravtsov2013, Hearin2019,Rodriguez2021}) and the so-called stellar-to-halo mass relation (SHMR, e.g., \citealt{Mandelbaum2006,Yang2009,Behroozi2010,Moster2010}).

The aforementioned scenario for galaxy formation takes place in an evolving Universe where structure grows hierarchically \citep{WhiteRees1978}. Smaller haloes merge to form larger structures, as dictated by the relentless pull of gravity. The existence of {\it{central}} and {\it{satellite}} galaxies in groups and clusters is therefore a natural consequence of the $\Lambda$CDM cosmological model. Central galaxies, which sit at the centre of big haloes, are larger and grow by absorbing material from satellites, which are in turn embedded in smaller subhaloes that are often in the process of accretion (e.g., \citealt{Larson1980, Balogh2000, Mccarthy2008, Campbell2015, Hearin2019, Jiang2020}).   

As satellite galaxies fall into groups and clusters, they are known to undergo a variety of (sometimes catastrophic) physical processes. By {\it{dynamical friction}} (e.g., \citealt{Chandrasekhar1943,BoylanKolchin2008}) energy can be transferred from the satellite's  orbit to the main halo, causing the former to spiral inwards. The strong tidal forces that satellites experience due to the presence of a central galaxy can remove DM, gas and stars from the galaxy, in a process called {\it{tidal stripping}} (e.g., \citealt{Merritt1983,Merritt1984,Merritt1985,Meyer2006,Read2006,Smith2016}).
This mechanism may also perturb the structure of galaxies, playing a significant role in galaxy evolution within clusters and groups of galaxies, changing the galaxies' morphologies and dynamics (e.g., \citealt{Valluri1993, Moore1999, Mastropietro2005}). In addition, if the halo contains a hot gas component, any gas associated with the satellite will experience a drag force due to the relative motion of the two fluids. If the drag force exceeds the restoring force due to the satellite’s own gravity, its gas will be ablated in a process called {\it{ram-pressure stripping}} (e.g., \citealt{Gunn1972,Abadi1999,Vollmer2001,Meyer2006, Safarzadeh2019, Boselli2022,RodriguezSilvio2022}). Other more subtle effects has been discovered, such as {\it{galaxy harassment}}, where galaxies are progressively ``heated" by high-speed encounters with other galaxies, becoming more prone to disruption by the potential well of the halo (e.g., \citealt{Moore1996,Moore1998, Gnedin2004, Smith2010}).

These processes have been invoked to explain morphological and other internal transformations of galaxies in high-density environments, giving rise, for instance, to the so-called morphology-density relation (\citealt{Dressler1980}). In ram-pressure stripping events, the decrease of the atomic and molecular gas components in the inner regions of the disk (the former partially swept away during the interaction, the latter
also gradually exhausted by the lack of feeding from the atomic gas reservoir), induces a decrease of the star formation activity (e.g., \citealt{Fillingham2016,Fillingham2018, Simpson2018}). This quenching process is indeed observed in the large
majority of HI-deficient galaxies in nearby clusters (e.g., \citealt{Boselli2014,Cybulski2014}). Under certain conditions, however, ram-pressure stripping can actually enhance activity, such as in the so-called jellyfish galaxies (e.g., \citealt{Vulcani2018,Safarzadeh2019}). 

\begin{figure*}
	\includegraphics[width=2.0\columnwidth]{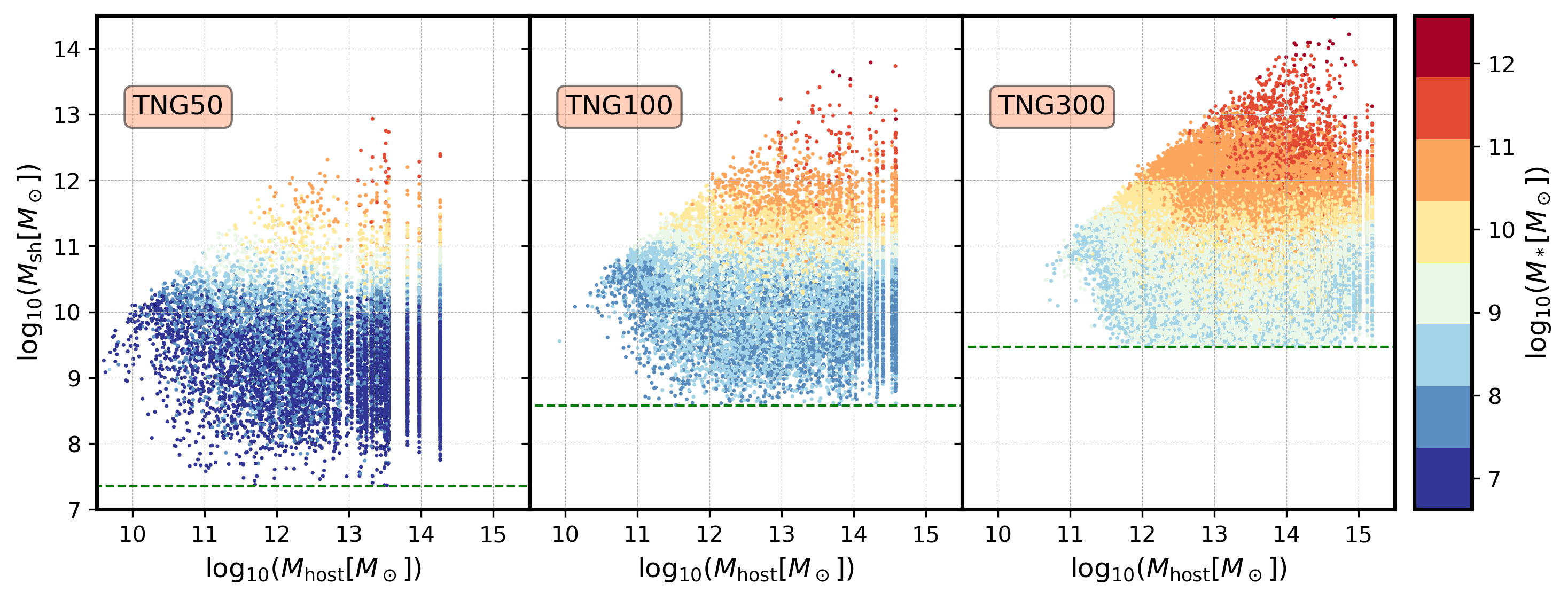}
    \caption{Total subhalo mass (which includes both baryons and DM) as a function of host halo mass for the 
    satellite galaxy population at $z=0$ in TNG50, TNG100, and TNG300, respectively. The colour code indicates the stellar mass of the corresponding satellite. The dashed lines show the resolution limits imposed on subhalo mass for each box. This figure illustrates the fact that each TNG box maps intrinsically different satellite galaxy populations. A complete study for a wide range of host and subhalo masses can only be performed when results from all three boxes are combined or discussed together.}
    \label{fig:scatter}
\end{figure*}

A particularly useful tool for studying the physical mechanisms involved in galaxy formation is provided by hydrodynamical simulations. These models employ known physics to simulate, at a sub-grid level, a variety of processes that are related to galaxy formation, including star formation, radiative metal cooling, and supernova, stellar, and black hole feedback (see reviews in \citealt{Somerville2015,Naab2017}). Nowadays, hydrodynamical simulations are used in a variety of astrophysical and even cosmological contexts. In this paper, we use the IllustrisTNG\footnote{\url{http://www.tng-project.org}} magnetohydrodynamical simulation suite (hereafter, simply TNG) to investigate and characterise the global effect that mass stripping has on the population of satellite galaxies, as predicted by the three main TNG boxes. To this end, we choose a population of galaxies that are accreted by a host halo at $z>1$ and maintain their identity as satellites all the way down to $z=0$. This restriction facilitates the analysis, while providing a long baseline to evaluate the average impact of these processes on the stellar mass, gas, and DM content of satellites, along with their half-mass radius and star formation properties. Our analysis also provides a useful comparison between boxes that serves as a benchmark for future analyses of satellite galaxies using TNG.

The effect of tidal stripping on satellite galaxies has been investigated before in TNG, from the perspective of the SHMR. \cite{Engler2021} showed that tidal stripping is responsible for the shift of the SHMR of satellites towards lower host halo DM masses, as compared to that of centrals. The reason is that tidal stripping occurs in an outside-in fashion, as previously shown by \cite{Smith2016}, so a large fraction of the subhalo is removed before the stellar mass component is significantly affected.  

The effect of tidal stripping has also been invoked to explain the possibility that galaxies without DM might exist, an aspect that we also address in this paper. By measuring the radial velocities of 10 luminous globular-cluster-like objects in the ultra-diffuse galaxy NGC1052-DF2, \cite{vD2018_noDM1} claimed the first detection of a galaxy only composed of baryonic matter. Soon after, a second candidate was discovered within the same galaxy group (NGC1052-DF4, \citealt{vD2019_noDM2}). Although a consensus on the nature of these systems has not been reached (it has been claimed that the galaxies could actually be at a much shorter distance, which would invalidate the conclusions of \citeauthor{vD2019_noDM2}), these results have stirred up discussion on the physical mechanisms that could produce these anomalies. \cite{Moreno2022}, in particular, using a hydrodynamical simulation, found that the tidal forces associated with extreme close encounters with massive neighbours can be responsible for the effect, predicting that $\sim$30$\%$ of massive central galaxies harbour at least one DM-deficient satellite. Also recently, \cite{vD2022} took a step forward suggesting that NGC1052-DF2 and NGC1052-DF4 are part of a trail of DM-free galaxies that originated from the same event, a collision of dwarf galaxies resembling the Bullet Cluster collision. In the last part of the paper, we apply the developed methodology to characterise this extreme galaxy population in TNG. 

This paper is organised as follows. Section \ref{sec:TNG} provides brief descriptions of the three TNG simulation boxes and the halo and galaxy properties analysed in this work. Section \ref{sec:evolution} addresses some important scaling relation for satellites, presenting a comparison of the halo mass -- galaxy size relations (\ref{sec:mass-size}) and the stellar-to-(sub)halo mass relations (\ref{sec:SsHMR}) measured in each box, including their redshift evolution. Section \ref{sec:stripping} shows our main results regarding the average effect of mass stripping on the satellite galaxy population from $z=1$, focusing both on the entire population and on low- and high-mass haloes separately (\ref{sec:combined}-\ref{sec:mhigh}). Also in Section \ref{sec:stripping}, we discuss the effect of resolution (\ref{sec:resolution}) and analyse the extreme case of DM-deficient galaxies (\ref{sec:DM}). Finally, Section~\ref{sec:discussion} is devoted to discussing the implications of our results and providing a brief summary of the paper. The TNG simulation adopts the standard $\Lambda$CDM cosmology \citep{Planck2016}, with parameters $\Omega_{\rm m} = 0.3089$,  $\Omega_{\rm b} = 0.0486$, $\Omega_\Lambda = 0.6911$, $H_0 = 100\,h\, {\rm km\, s^{-1}Mpc^{-1}}$ with $h=0.6774$, $\sigma_8 = 0.8159$, and $n_s = 0.9667$.

\section{The Illustris TNG hydrodynamical simulation}\label{sec:TNG} 

In this work, we analyse the satellite galaxy population using the three main boxes of the TNG suite: TNG50-1, TNG100-1, and TNG300-1 \citep[hereafter TNG50, TNG100, and TNG300, respectively, see][]{Pillepich2018b,Pillepich2018,Nelson2018_ColorBim,Nelson2019,Marinacci2018,Springel2018}. The TNG simulations were carried out using the {\sc arepo} moving-mesh code \citep{Springel2010} and feature sub-grid models that account for a variety of physical processes (e.g., star formation, stellar/supermassive black hole feedback, chemical enrichment from SNII, SNIa, and AGB stars, and radiative metal cooling). Although the TNG boxes span different sizes and resolutions, they all share the same sub-grid physics model, except for the equation of the state of star-forming gas that was updated in TNG50. This modification, however, has no impact on the galactic properties, as shown in \citet{Nelson2019}.

TNG300 is the largest box of the suite, with a side length of $205\,\,h^{-1}$Mpc. It also has the lowest DM resolution, with a DM particle mass of 5.9$\times 10^{7}$ M$_\odot$. TNG100 has a side length of $75\,\,h^{-1}$Mpc, with a factor of 8 higher resolution than TNG300 (7.5$\times 10^{6}$ M$_\odot$ for DM). Finally, TNG50 is the smallest box ($35\,\,h^{-1}$Mpc) and also the better resolved (4.5$\times 10^{5}$ M$_\odot$ for DM, a factor 130 better than TNG300). The TNG suite has shown remarkable agreement with several observational results at different redshifts, proving to be a self-consistent and powerful tool to investigate galaxy formation and the halo-galaxy connection (see, e.g., \citealt{Springel2018,Pillepich2018,Bose2019,Beltz-Mohrmann2020,Contreras2020,Gu2020,Hadzhiyska2020,Hadzhiyska2021,Shi2020,MonteroDorta2020A,MonteroDorta2021A,MonteroDorta2021C,Rodriguez2021,Favole2022}, to name but a few).

DM haloes in TNG are identified using a friends-of-friends ({\sc fof}) algorithm \citep{Davis1985}, while substructure (i.e., subhaloes) are identified using the {\sc subfind} algorithm \citep{Springel2001}. Subhaloes containing a non-zero stellar component are considered galaxies. Importantly, this work makes use of the merger trees of each simulated box, computed using {\sc sublink} \citep{Rodriguez-Gomez2015}, and available in the TNG database. These merger trees are a key piece of our analysis, allowing us to map the evolution of individual satellite galaxies across cosmic time. Here we use the main branch of subhaloes between $z=0$ and $z=2$ (even though most of the analysis is restricted to $0<z<1$).

Several subhalo and halo properties available in the TNG database are employed in this work. For DM haloes, we use the virial mass, $M_{\rm host} [{\rm M_\odot}]$, defined as the total mass enclosed within a radius where the density equals 200 times the critical density (this includes both baryons and DM). For galaxies, we analyse the stellar mass, $M_{*} [{\rm M_{\odot}}]$, gas mass, $M_{\rm gas} [{\rm M_{\odot}}]$, and DM mass, $M_{\rm DM} [{\rm M_{\odot}}]$, computed respectively, as the total mass of stellar particles, gas cells, and DM particles bound to each subhalo. We also employ the total mass of each subhalo, $M_{\rm sh} [{\rm M_{\odot}}]$, where all components are considered. Other than the mass components, we use the star formation rate ($SFR[{\rm M_{\odot} yr^{-1}}]$), and the galaxy half-mass radius ($R_*[{\rm kpc}]$). For the latter, we use both the three-dimensional half-mass radius provided by the simulation and the galaxy size data from \citet{Genel2018}.

Based on the resolution, several mass cuts are imposed on each box. First, only subhaloes with stellar masses at $z=0$ greater than 50 initial gas cells are considered. This yields minimum stellar masses of $4.25\times10^{6}\,\, {\rm M_{\odot}}$, $7.00\times10^{7}\, \, {\rm M_{\odot}}$, and $5.50\times10^{8}\,\, {\rm M_{\odot}}$ for TNG50, TNG100, and TNG300, respectively. A similar resolution threshold is chosen for subhalo mass, $M_{\rm sh}$, which includes all material components. We impose a threshold equivalent to 50 DM particles, resulting in  
minimum subhalo masses of $2.25\times10^{7}\,\,{\rm M_{\odot}}$, $3.75\times10^{8}\,\,{\rm M_{\odot}}$, and $2.95\times10^{9}\,\,{\rm M_{\odot}}$, respectively.

The satellite galaxy populations mapped by the three TNG boxes at $z=0$ after the aforementioned cuts are applied are compared in Fig. \ref{fig:scatter}. For each of them, the total subhalo mass is shown as a function of host halo mass, with a colour code indicating the stellar mass of the hosted satellite. This figure illustrates an obvious difference between boxes that the reader should bear in mind throughout this work: the typical mass of subhaloes (including stellar, gas, and DM mass) increases significantly with box size in TNG. Fig. \ref{fig:scatter} also illustrates an obstacle for the analysis, i.e., the fact that the selections in the TNG boxes display only mild overlap. This limitation hinders any analysis performed at fixed stellar and subhalo DM mass, while maintaining good statistics (recall that we are only interested in satellites that remain satellites from $z=1$).

\begin{figure}
    \includegraphics[width=1.0\columnwidth]{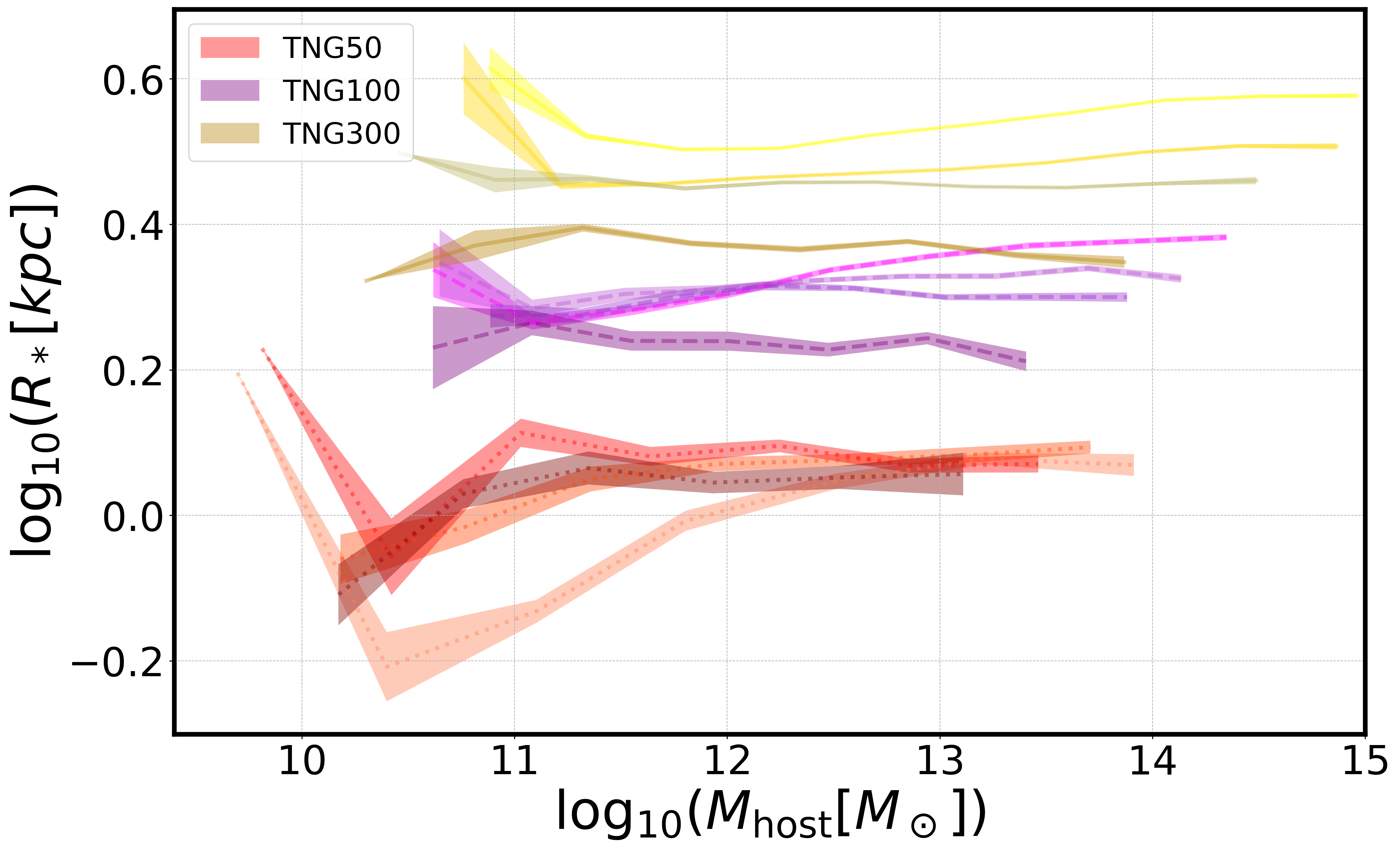}
    \caption{The host halo mass -- galaxy size relation for satellite galaxies at different redshifts. The size measurements correspond to the 3D half-mass radii from Genel et al. (2018). Results from TNG50, TNG100, and TNG300 are shown in red, purple, and yellow, respectively, for the redshift snapshots $z=0,0.5,1.0$ and 2 (colours are darker as redshift increases). The shaded regions indicate the standard error on the mean.}
    \label{fig:SHM-evo}
\end{figure}

\begin{figure}
     \includegraphics[width=0.9\columnwidth]{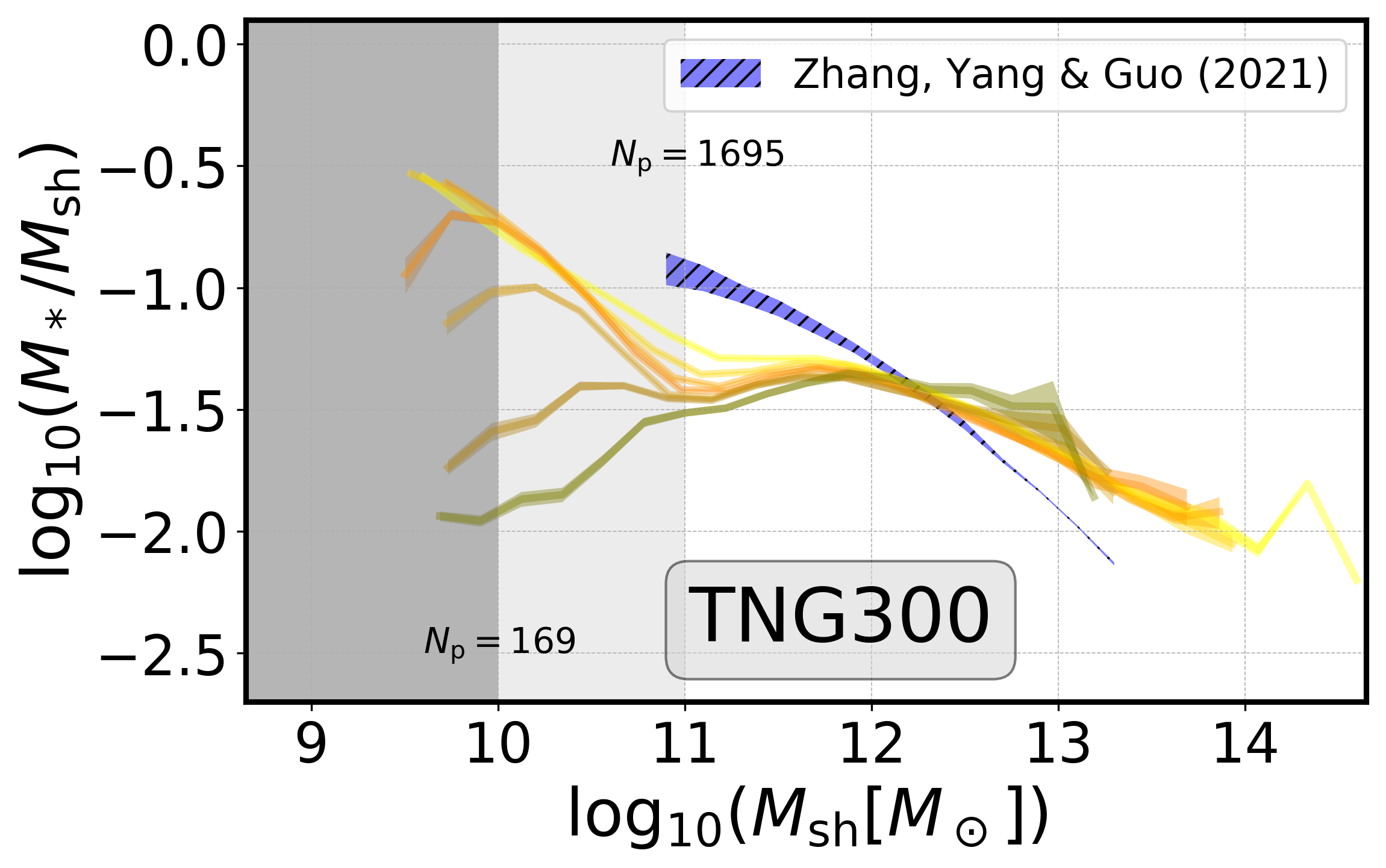}
     \includegraphics[width=0.9\columnwidth]{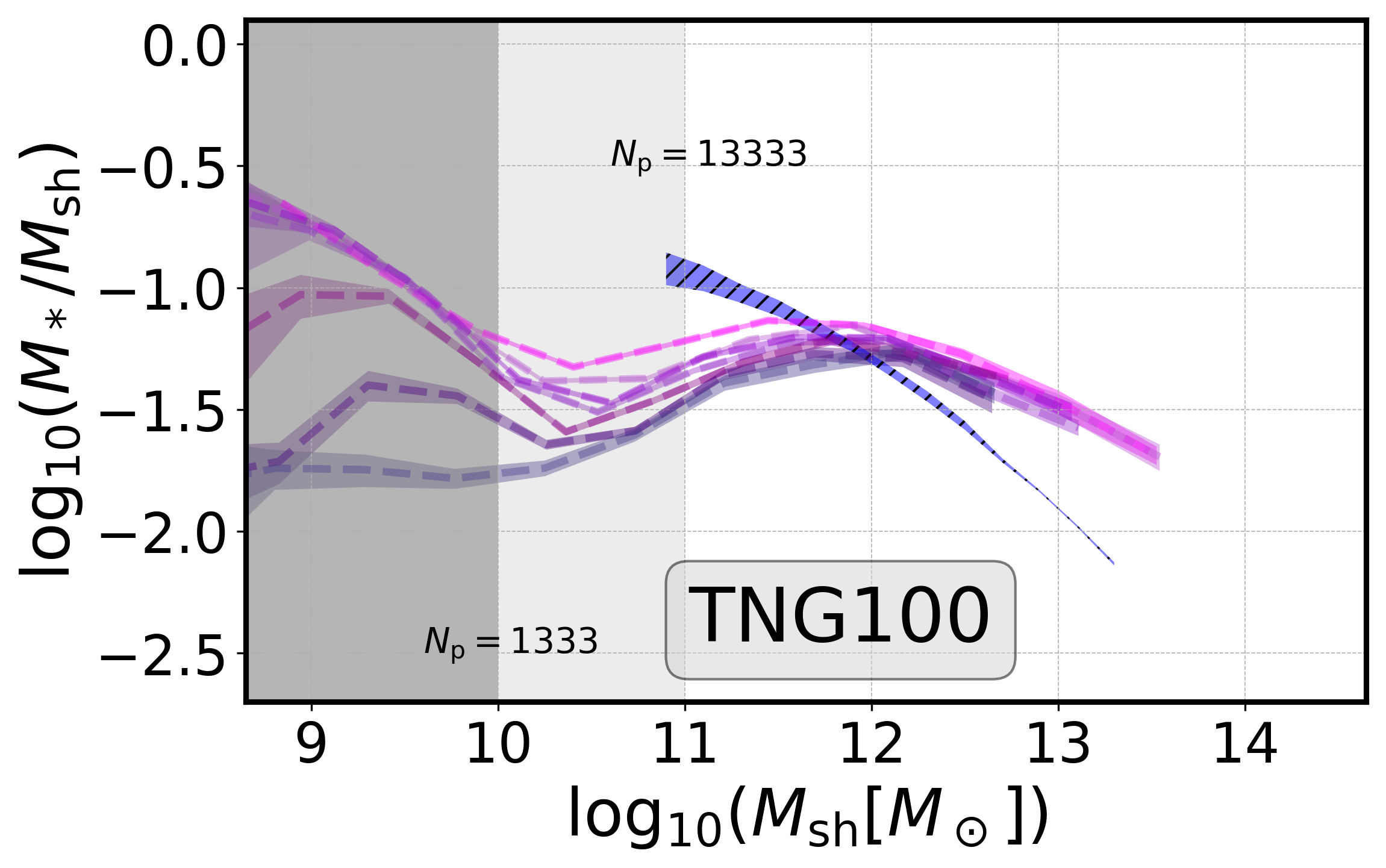}
     \includegraphics[width=0.9\columnwidth]{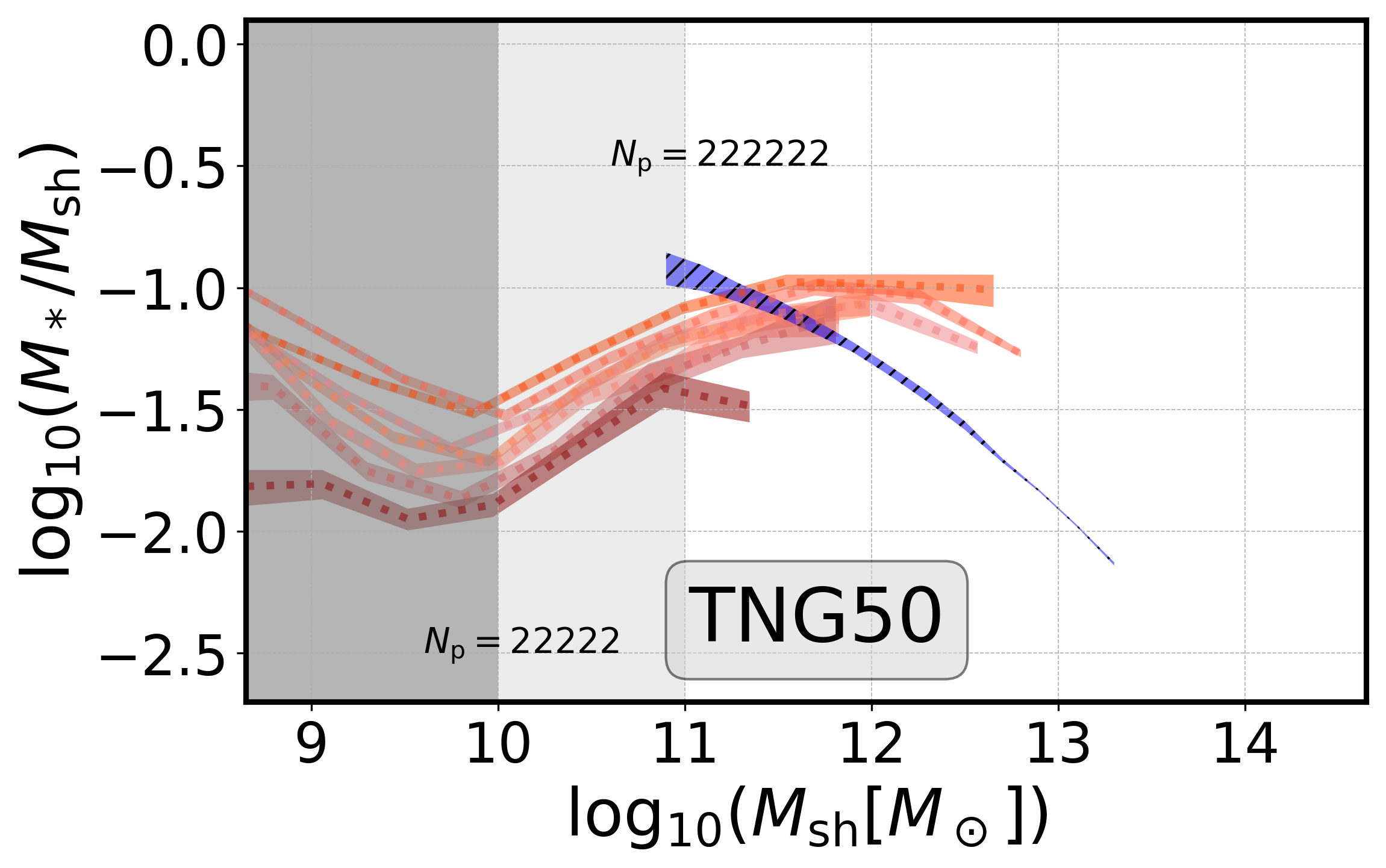}

    \caption{Redshift evolution of the stellar-to-(sub)halo mass relation (SsHMR) for satellite galaxies in TNG. Results from TNG50, TNG100, and TNG300 are shown in different panels for the redshift snapshots $z=0,0.5,1.0$ and 2 (colours are darker as redshift increases). The shaded regions around the curves indicate the standard error on the mean, whereas the vertical grey regions show the subhalo mass ranges where statistics could be potentially problematic (according to previous works). The number of DM particles, $N_{\rm p}$, corresponding to $\log_{10}M_{\rm host}[{\rm M_\odot}]=10$ and 11 are included for reference. In blue, results obtained by \citealt{Zhang2021} at $z=0$ are also plotted for comparison (these results were obtained with a different data set, see text).}
    \label{fig:SHMR-evo}
\end{figure}

It is pertinent to mention a potential caveat of our analysis. Spherical overdensity finders such as the {\sc subfind} algorithm are known to suffer from identification issues when substructures are too close. This could happen, for instance, when satellites are adjacent to the pericenter of a larger halo \citep{Muldrew2011,Springel2021}. We palliate this by taking (population) averages and analysing the dependence of satellite properties on halo-centric distance for different distance bins. Note also that using merger trees is helpful in this context: even though {\sc subfind} might miss a particular subhalo at a certain snapshot,  the merger tree algorithm will search for this subhalo at the next snapshot. We have also checked that removing objects with very small halo-centric distances (e.g., $\lesssim 0.03R_{\rm vir}$) has almost no quantitative impact on our results. 

Another aspect that must be carefully considered in this type of studies is the integration timescale of numerical simulations. Long temporal spacing between consecutive snapshots could result in spurious features in the geometry of galaxy interactions, missed pericenters,  and underestimation of star formation \citep[see e.g.,][]{Ploeckinger2018,Rodriguez2021}. IllustrisTNG uses a highly adaptive time and space integration scheme where cells and particles evolve on individual time-steps \citep[][]{Pillepich2018b}. In this work, we cover an age range of $\sim 10$ Gyr ($z=0-2$) and the IllustrisTNG snapshots are temporally spaced by $\sim 100-200$~Myr. For this type of spacing, missing pericenters is unlikely (the pericenter is the minimum-distance passage between satellites and centrals, as discussed below). 

\section{The evolution of scaling relations for satellite galaxies}
\label{sec:evolution} 

Galaxies form and evolve inside DM haloes, which establishes an intrinsic connection that manifests itself in several halo and galaxy properties. In this section, we contextualise our analysis by presenting the evolution of satellite galaxies from the perspective of their {\it{scaling relations}}.

\subsection{Halo mass -- galaxy size relation}
\label{sec:mass-size}

Among the main properties of galaxies, the half-mass radius provides valuable information on the galaxy formation and evolution process, while DM halo mass is linked to the assembly history of their hosting haloes. The halo mass -- galaxy size relation was investigated for central and satellite galaxies separately in \cite{Rodriguez2021}. At redshift $z=0$, the results reported in that work indicate that, both in observations from the Sloan Digital Sky Survey Data Release 7 \citep[SDSS DR7;][]{Abazajian2009} and simulations from TNG300, the size of central galaxies increases with halo mass, while the mean size of satellite galaxies depends only slightly on the mass of their host halo. In this section, we extend this analysis by addressing the redshift evolution of the halo mass -- galaxy size relation for satellites in TNG50, TNG100, and TNG300. Our analysis  provides insight not only on the physical processes that shape the evolution of satellites, but also on the effects of resolution and volume on the measurements. 

In Fig. \ref{fig:SHM-evo}, the halo mass -- galaxy size relation is displayed for satellite galaxies at four different redshifts ($z=0.0,0.5,1.0,2.0$, respectively) for TNG50, TNG100, and TNG300. In order to follow up on the \cite{Rodriguez2021} work, we have employed the size determinations from \cite{Genel2018}, which are available for the aforementioned redshifts. In particular, the relation is measured using the 3D half-mass radius. Details on the measurements and selection criteria, which dictate the lower halo mass limits in Fig. \ref{fig:SHM-evo}, can be found in \cite{Genel2018}\footnote{We have checked that the mass --size relation does not change significantly when the standard TNG size measurements are employed.}. Generally speaking, Fig. \ref{fig:SHM-evo} confirms the results reported in \cite{Rodriguez2021} at $z=0$, showing little dependence of the size of satellites on the mass of their host haloes, even when different redshifts and boxes are considered. Some exceptions to this general trend are, however, noticeable. For a given box, the most relevant deviation is apparent for TNG50 at low halo masses, where the average half-mass radius drops significantly at all redshifts (at very small masses it goes up again). This seemingly erratic behaviour is relatively surprising, since TNG50 is by construction better equipped to deal with small haloes. Note that these are all haloes that are above reasonable resolution thresholds.

Clearly more significant than the above deviations is the dependence of the average size of satellites on the box employed, and, thus, on resolution and volume. As resolution decreases with volume, the average size of TNG galaxies increases, with differences between boxes of the order of $\sim 0.2$~dex. This behaviour was expected from the results shown in Fig. \ref{fig:scatter}. It has also been documented before in the literature  \citep[stellar mass from TNG300 and TNG100 is typically rescaled by a factor of $\sim 2$, and $\sim1.5$, respectively, see e.g.][]{Vogelsberger2018,Vogelsberger2020,Engler2021}. Continuing with the discussion on the mean size of satellites, the redshift evolution predicted by different boxes seems to be slightly different as well. While in TNG300 satellites are on average smaller at higher redshift, this effect is not as clear in TNG50 and TNG100. Again, these different behaviours are seemingly related to the different populations mapped by the boxes.  

Despite the aforementioned differences, there is not a strong case for a dependence of galaxy size on host halo mass for satellites, independently of the redshift considered. This result is not obvious. On the one hand, more massive haloes could be more likely to accrete larger galaxies, which could potentially imprint a dependence on halo mass. Of course, the picture is complicated by the physical processes that take place inside haloes. Mechanisms such as mass stripping play a key role in the dynamical evolution of satellite galaxies. Despite this complexity, the absence of any significant correlation between their sizes and the properties of their hosting haloes is an important result.

Finally, we have checked that the differences between boxes manifest themselves also at fixed stellar mass. The stellar mass-size relation displays a higher amplitude in larger boxes (with variations that depend on stellar mass). At fixed stellar mass, there is a higher fraction of large galaxies in TNG300 as compared to TNG50. This might in principle have an impact on the amount of mass stripping predicted by each box (more extended disks could be more sensitive to stripping). As we will show in the following sections, this effect is, however, not noticeable in the mean evolution of satellite galaxies.

\begin{figure*}
    \includegraphics[width=2.05\columnwidth]{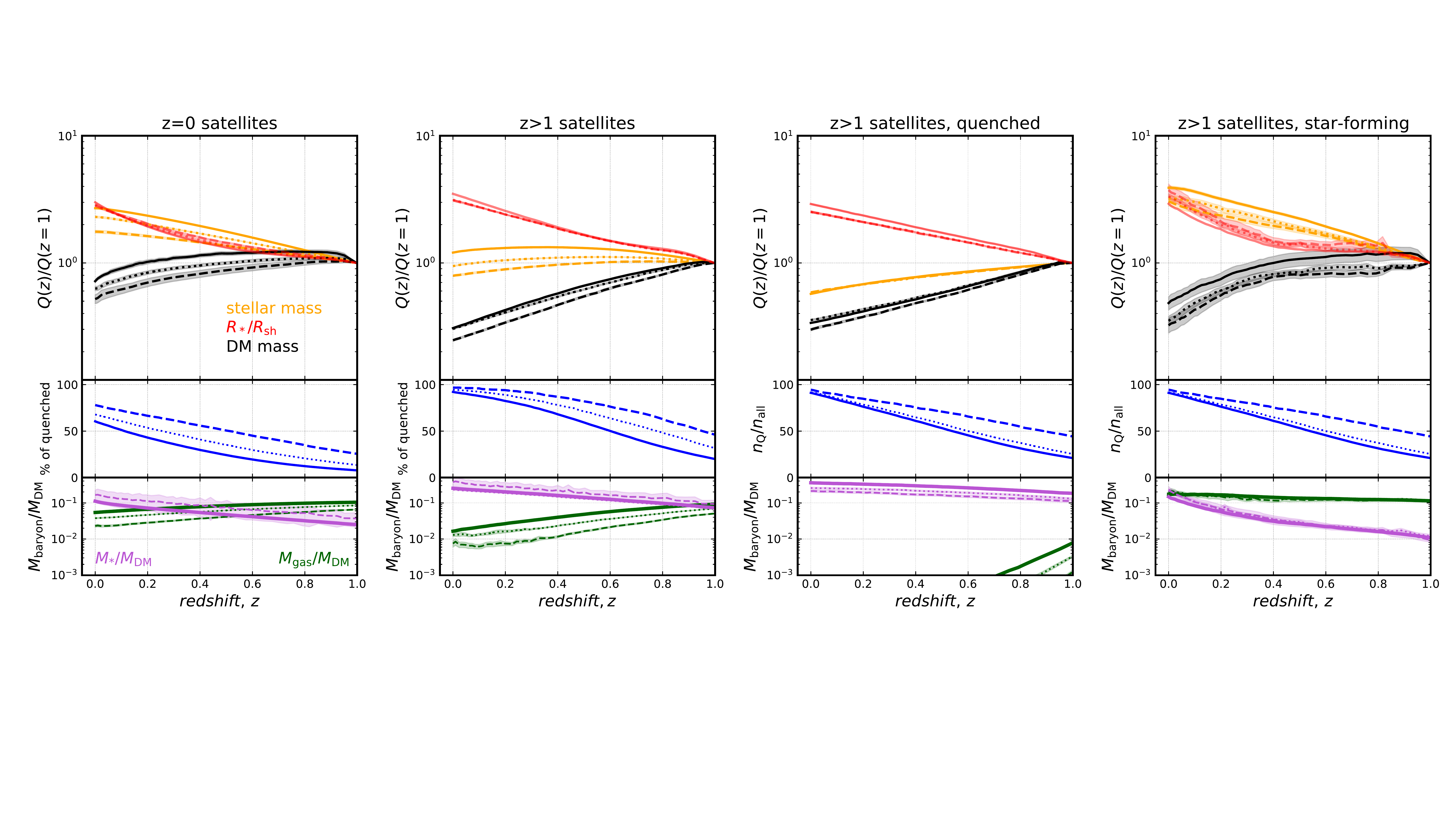}
    \caption{The redshift evolution of some of the main properties of satellite galaxies in different subsets. From left to right: all satellites passing our resolution cuts at $z=0$, our fiducial $z>1$ sample (satellites accreted at $z>1$ that maintain their identity all the way through), quenched $z>1$ satellites, and star-forming $z>1$ satellites (here, we include satellites that remain quenched/star-forming during the entire redshift range, see text). {\it{Top panels:}} The redshift evolution of the stellar and DM content of subhaloes, along with the stellar-to-subhalo half-mass radius ratio ($R_{*}/R_{\rm sh}$). Each curve is normalised to the corresponding value at $z=1$. Results from the TNG50, TNG100, and TNG300 boxes are represented by dashed, dotted, and solid lines, respectively. {\it{Middel panels:}} In the two left-hand columns, the quenched fraction for each subset (i.e., the fraction of quiescent satellites of the entire satellite population considered) is displayed as a function of redshift. In the two right-hand columns, we show the fraction of the entire population at each redshift that is quenched or star-forming, respectively. {\it{Bottom panels:}} The average gas-to-DM and stellar-to-DM mass fractions for each population and box considered. The uncertainties on the measurements, represented by shaded regions, correspond to the statistical errors on the mean.}
    \label{fig:sats_entire}
\end{figure*}

\subsection{Stellar-to-(sub)halo mass relation}
\label{sec:SsHMR}

There is a tight physical connection between the properties of satellite galaxies and the properties of their subhaloes. In this section, we explore this connection by measuring the evolution of the stellar-to-(sub)halo mass relation (SsHMR) for satellite galaxies. These results (i.e., the ratio of stellar mass to subhalo mass as a function of subhalo mass) are presented in Fig. \ref{fig:SHMR-evo} for the three TNG boxes. Note that subhalo mass includes all material components. 

The intrinsic differences in the population of subhaloes mapped by each box have again an impact on the SsHMR measurement, despite a general, qualitative agreement. For all boxes, haloes above a certain mass seem to display a relation that resembles in shape that measured for central galaxies \citep[see, e.g.,][]{Behroozi2010}, including the characteristic peak of galaxy formation efficiency. This is in agreement with previous findings from \cite{Engler2021}. The subhalo mass of the peak seems consistent for different boxes and redshifts, although the peak mass fraction is larger for the smaller boxes (note that TNG50 does not reach high host halo masses with enough statistics, so the subsequent drop is not observed). The characteristic mass above which the SsHMR behaves like that of centrals varies from box to box, ranging from $\log_{10}M_{\rm host}[{\rm M_\odot}]\simeq 11$ in TNG300 to $\simeq$10 in TNG50 (with some redshift dependence). In Fig. \ref{fig:SHMR-evo}, we show the number of DM particles that correspond to these thresholds.  

Below the characteristic mass, and particularly for low redshifts, a significant enhancement in the fraction is observed. This increase in the stellar mass of subhaloes is attenuated as we move to higher redshifts; at $z=2$, no enhancement is observed. One possibility is that the effect is completely unphysical and caused by resolution issues (in \citealt{Engler2021}, the range of subhalo masses is restricted to $\log_{10}M_{\rm host}[{\rm M_\odot}]\gtrsim 11$). At faced value, however, there are physical effects that could explain the trends. First, we will show in following sections that lower-mass subhaloes (particularly in low-mass host haloes) lose DM material more easily due to mass stripping. Stripping would also affect the stellar component, but in a lesser extent since this is more concentrated inside the potential well of the subhalo (e.g., \citealt{Smith2016}). This would explain the fact that the enhancement becomes stronger at low redshift (since subhaloes would have been exposed to mass stripping for a longer period of time). Second, the resolution cuts that we impose, which run diagonally on the $M_*/M_{\rm sh}$ vs. $M_{\rm sh}$ plane, would tend to amplify the effect on the average $M_*/M_{\rm sh}$ fraction, perhaps artificially. This, in combination with the resolution itself, would explain the fact that the subhalo mass at which the enhancement becomes significant decreases for smaller boxes. It is important to stress that the low-mass enhancement occurs in TNG50 at subhalo masses that appear to be sufficiently well resolved (e.g., $\simeq 7000$ DM particles at $\log_{10}(M_{\rm sh}[{\rm M_\odot}])=9.5$). We have also checked that a more restrictive stellar mass resolution cut does not eliminate the feature.

There is little discussion on the SsHMR in the literature. In \citet{Engler2021}, the relation is measured and discussed in detail for the TNG boxes for subhaloes above $\log_{10}(M_{\rm sh}[{\rm M_\odot}])=10.5-11$. Importantly, they compare the SsHMR of satellites and centrals, showing that, at fixed stellar mass, the former is shifted towards lower dynamical masses. This effect is attributed, again, to tidal stripping of the DM component working outside-in inside the subhalo \citep{Smith2016}. In Fig. \ref{fig:SHMR-evo}, we explicitly compare our results with those presented in \cite{Zhang2021}, where galaxies from the SDSS are linked to subhaloes from the ELUCID simulation. The SsHMR from \cite{Zhang2021} decreases with host halo mass from 
$\log_{10}(M_{\rm sh}[{\rm M_\odot}])=11$, the lowest halo mass considered. Unlike for TNG, there is no peak or plateau in the SsHMR of \cite{Zhang2021} within the halo mass range considered.

The scaling relations presented for satellite galaxies reflect the intrinsic link between the baryonic and DM components of haloes in the TNG boxes. In the next section, we will address in more depth the individual evolution of subhaloes harbouring satellite galaxies to evaluate the effect that mass stripping has on their material components. To avoid potential resolution issues, we will restrict the analysis to satellites in host haloes with masses above $\log_{10}( M_{\rm host}[{\rm M_\odot}])=11$.

\section{The evolution of satellites in TNG}
\label{sec:stripping}

\begin{figure*}
	\includegraphics[width=2.0\columnwidth]{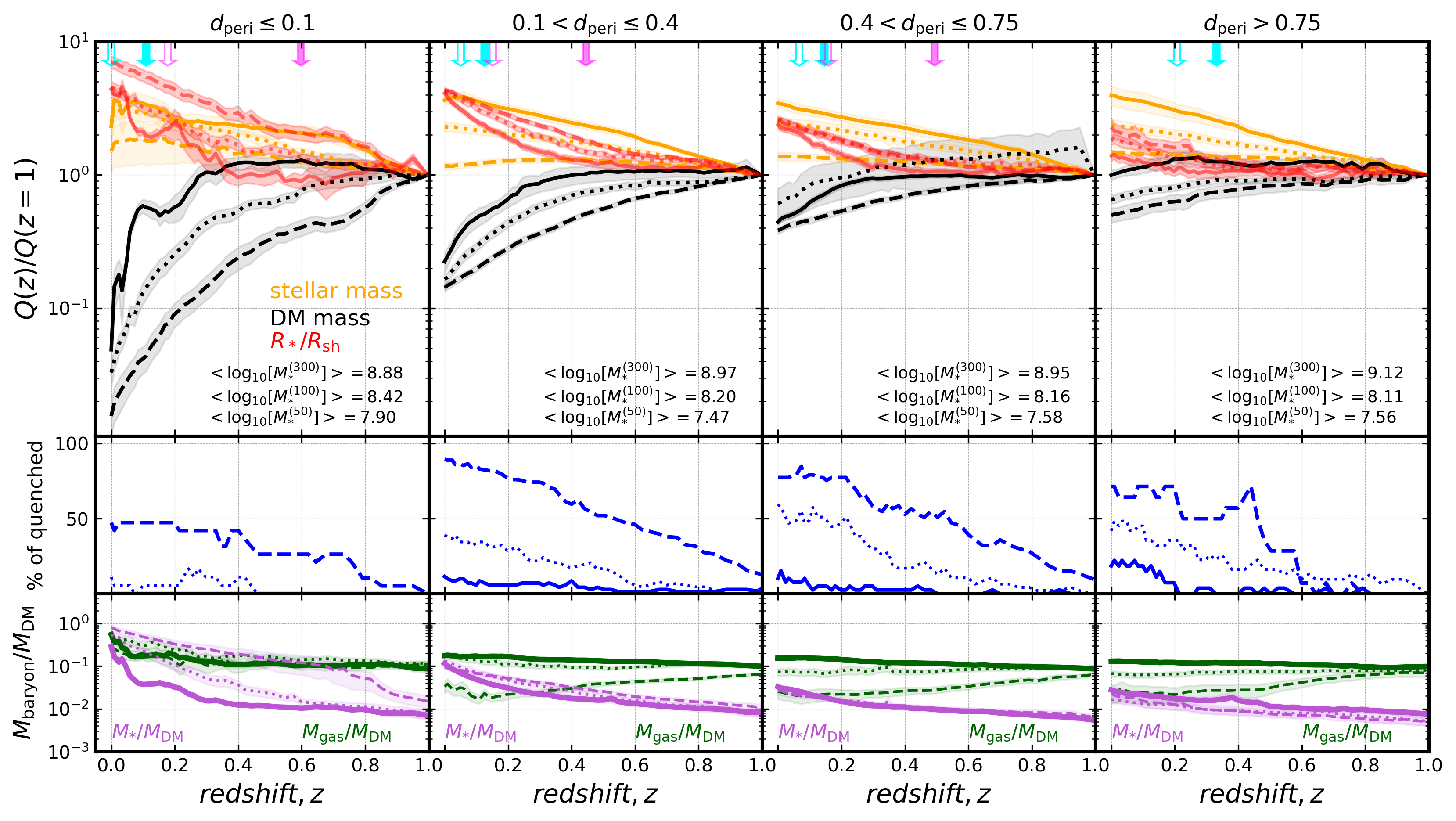}
    \caption{The same as Fig. \ref{fig:sats_entire} but for $z>1$ satellites (fiducial sample) pertaining to host haloes of mass $11<\log_{10} (M_{\rm host}[{\rm M_\odot}])<12$. From left to right, each plot corresponds to satellites with increasing pericentric distance (see text and top label).  We have also added several arrows indicating: 1) the average pericentric redshift, $z_{\rm peri}$, in the subsets (i.e., the average redshift at which galaxies have their pericentric passage, cyan arrows), 2) the average redshift at which the population reaches within a certain distance to the centre for the first time (magenta arrows). This reference distance is set at 0.4$R_{\rm vir}$ in the first and second panels (from left to right) and 0.75$R_{\rm vir}$ in the third panel. Solid arrows correspond to the TNG50 box, whereas open arrows show results for TNG300. Magenta arrows are not shown for the last $d_{\rm peri}$ bin since satellites remain far from the halo centre at all times. Finally, the average stellar masses for the subsets are indicated in the legend.}
    \label{fig:sats_m1}
\end{figure*}

\subsection{The combined satellite population}
\label{sec:combined}

As galaxies fall into groups and clusters, they experience a variety of physical processes that can alter their structure and matter content. In particular, processes such as tidal stripping result in varying degrees of mass loss, including stellar, DM, and gas mass. It is also well documented that both tidal and ram-pressure stripping can produce the reduction and even truncation of star formation (e.g., \citealt{Fillingham2016,Fillingham2018, Simpson2018}). One of the main goals of this paper is to quantify the combined effect of different mass-stripping mechanisms in the TNG boxes. It is important to emphasise again that the TNG boxes map intrinsically different satellite galaxy populations due to the differences in box size and resolution (see Fig. \ref{fig:scatter}). Discussing results from different boxes is therefore the best strategy to investigate the evolution of this galaxy population. In order to ensure some mass resolution at higher redshifts, we have imposed an additional stellar mass cut: $M_* > 10^7 \, {\rm M_\odot}$ for TNG50 and TNG100, and $M_* > 10^8 \, {\rm M_\odot}$ for the lower resolution TNG300, across all snapshots. 

We have opted to analyse only galaxies that are accreted at $z>1$ and maintain their identity as satellites all the way down to $z=0$. The motivation for this choice is twofold. First, it facilitates the analysis of the evolution of objects along the merger tree. Second, it allows us to probe a population of satellites that have been subject to mass stripping for a significant period of time. Note that our results are qualitatively consistent when the satellite redshift condition is shifted to $z=0.5$ and $z=2$. We have also checked that the redshift of last infall for these satellites is quite consistent between boxes: $1.57 \pm 0.35$, $1.53 \pm 0.34$, and $1.48 \pm 0.33$ for TNG50, TNG100, and TNG300, respectively (where the variation ranges are simply obtained from the standard deviations). Here, the infall redshift is simply defined as the redshift at which galaxies become satellites for the last time (i.e., the last transition from central to satellite). 

\begin{figure*}
	\includegraphics[width=2.0\columnwidth]{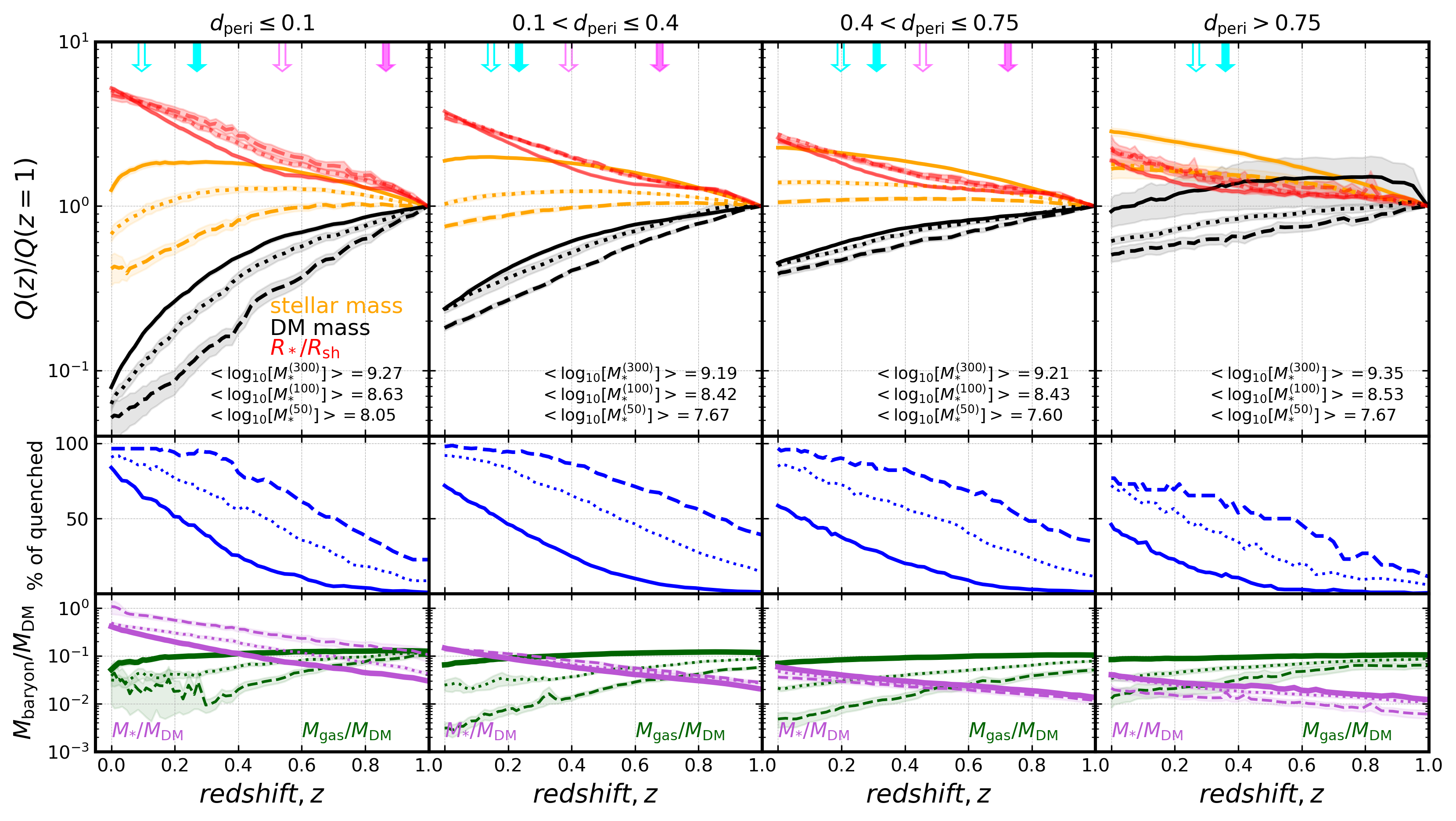}
    \caption{The same as Fig. \ref{fig:sats_m1} but for $z>1$ satellites pertaining to host haloes of mass $12<\log_{10}(M_{\rm host}[{\rm M_\odot}])<13$.}
    \label{fig:sats_m2}
\end{figure*}

\begin{figure*}
	\includegraphics[width=2.0\columnwidth]{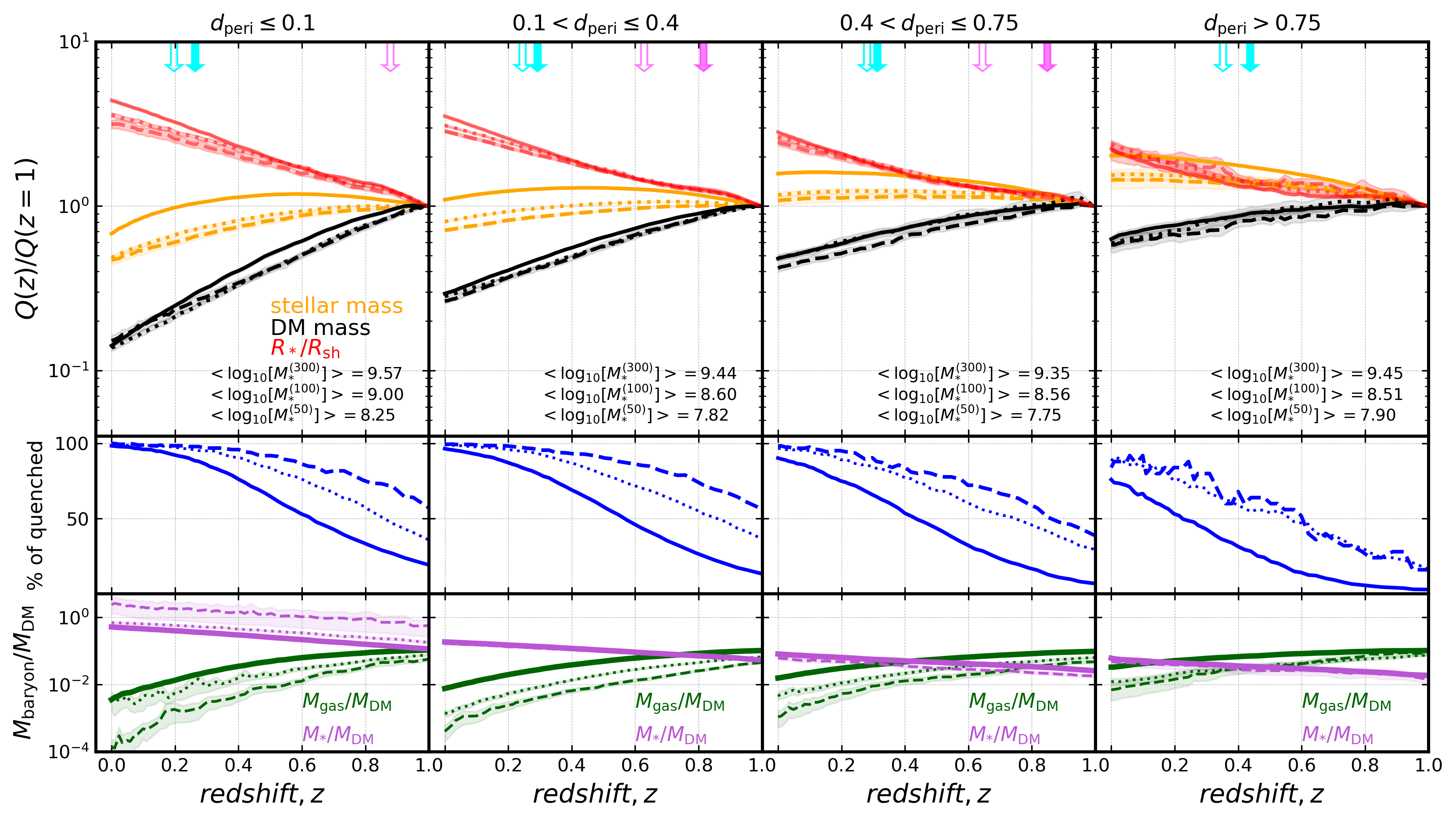}
    \caption{The same as Fig. \ref{fig:sats_m1} but for $z>1$ satellites pertaining to host haloes of mass $13<\log_{10}(M_{\rm host}[{\rm M_\odot}])<14$.}
    \label{fig:sats_m3}
\end{figure*}

\begin{figure*}
	\includegraphics[width=2.0\columnwidth]{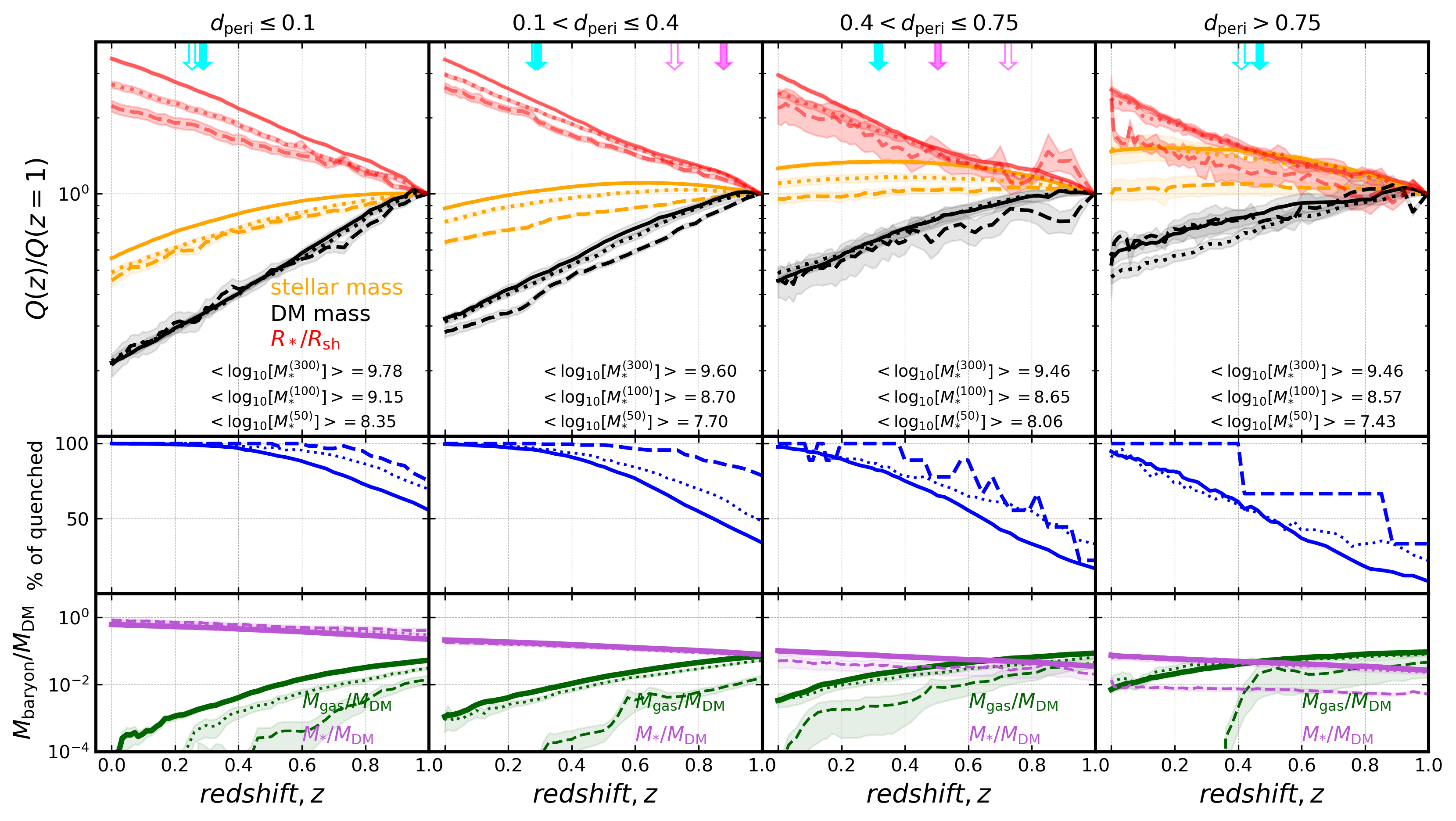}
    \caption{The same as Fig. \ref{fig:sats_m1} but for $z>1$ satellites pertaining to host haloes of mass $14<\log_{10}(M_{\rm host}[{\rm M_\odot}])<15$.}
    \label{fig:sats_m4}
\end{figure*}

\begin{figure}
	\includegraphics[width=1.0\columnwidth]{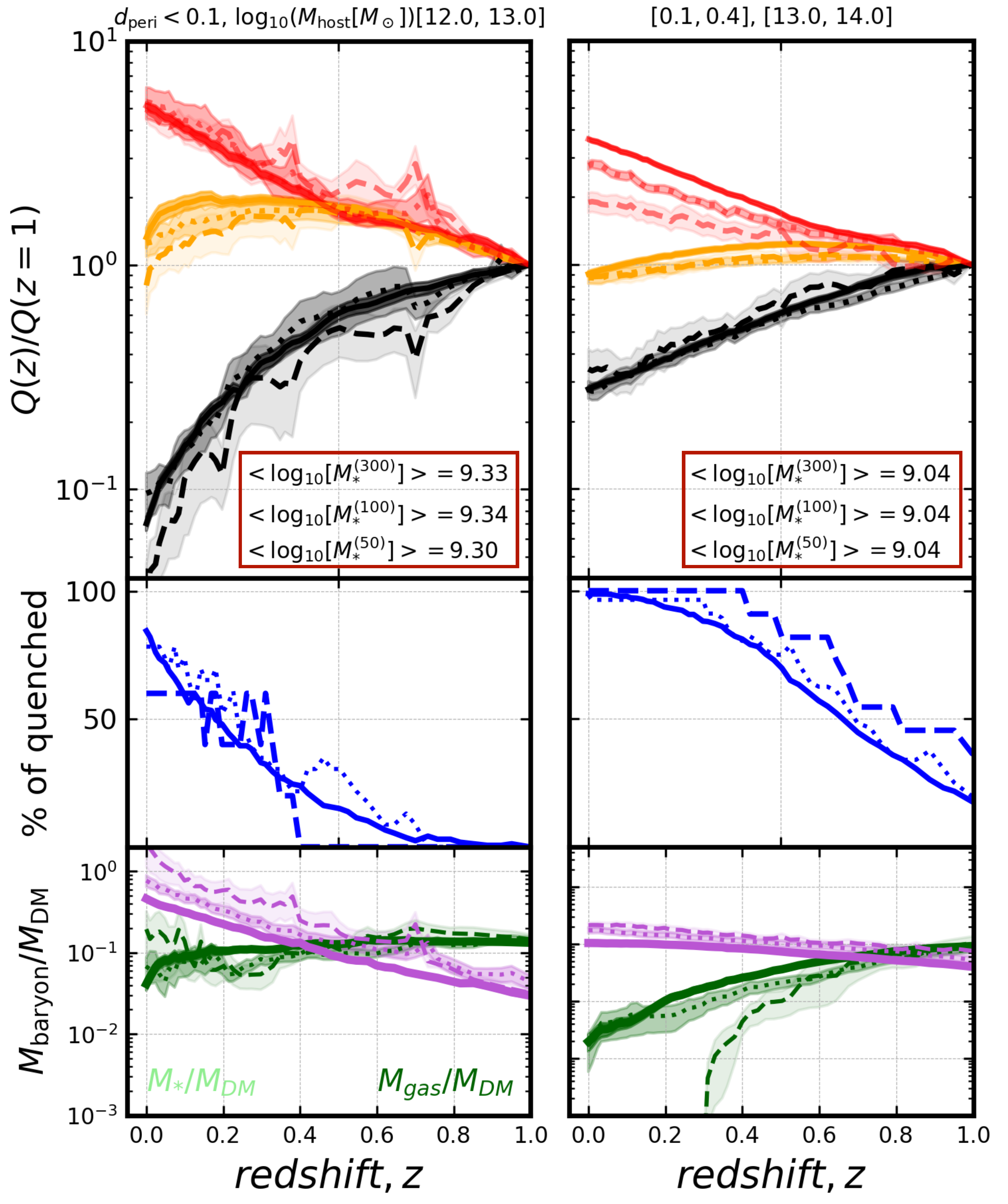}
    \caption{The effect of fixing stellar mass in the redshift evolution plots for two bins in halo mass -- $d_{\rm peri}$ space for which statistics are good enough. {\it{Left-hand plot}}: redshift evolution, in the same format employed before, for satellites in host haloes of mass $12<\log_{10}(M_{\rm host}[{\rm M_\odot}])<13$ and pericentric distance $d_{\rm peri} \le 0.1$.  {\it{Right-hand plot:}} redshift evolution for satellites in host haloes of mass $13<\log_{10}(M_{\rm host}[{\rm M_\odot}])<14$ and pericentric distance $0.1<d_{\rm peri} \le 0.4$. The uncertainties on the measurements, represented by shaded regions, correspond to the statistical errors on the mean.}
    \label{fig:tests}
\end{figure}

Fig. \ref{fig:sats_entire} compares the 
redshift evolution of the aforementioned population (second panel, from left to right) with that of $z=0$ satellites (i.e., imposing the ``satellite" status only at $z=0$, first panel). The average infall redshift for this population in the boxes is $\sim 0.7$, with a standard deviation of $\sim 0.6$. The format displayed in Fig. \ref{fig:sats_entire} will be replicated in figures \ref{fig:sats_m1} -  \ref{fig:tests}. The uncertainties on the measurements, represented by shaded regions, correspond to the standard errors on the mean (the object-to-object variation in these evolutionary trends for galaxies is known to be quite large). It is noteworthy that the results presented in Fig. \ref{fig:sats_entire} are qualitatively consistent with those from \cite{ChavesMontero2016}, using the EAGLE simulation (see their Fig. 12). 

We begin by analysing the $z=0$-selected satellite population (leftmost panel of Fig. \ref{fig:sats_entire}). The upper panel displays the average evolution of stellar and DM subhalo mass, along with the stellar-to-subhalo size ratio (i.e., the galaxy half-mass radius divided by subhalo half-mass radius, $R_*/R_{\rm sh}$), normalised to the corresponding value at $z=1$. 
Here, the TNG50, TNG100, and TNG300 results are represented by dashed, dotted, and solid lines, respectively. For statistical reasons, we have opted to employ the size measurements provided in the simulations, instead of those from \cite{Genel2018}, but we have checked that our results do not depend strongly on this choice. Fig. \ref{fig:sats_entire} shows that, qualitatively, all boxes provide similar predictions. The average stellar mass grows from $z=1$ to $z=0$, whereas DM mass experiences a modest decrease due to tidal stripping (less than $50\%$ loss). As a consequence, the size ratio increases significantly, more rapidly than stellar mass. This picture is in agreement with previous findings that tidal stripping proceeds in an outside-in fashion: the stellar component, which is more concentrated and thus more tightly bounded to the potential well of the halo, is less affected by tidal effects (see, \citealt{Engler2021, Smith2016}).

The middle panel (of the leftmost plot in Fig. \ref{fig:sats_entire}) shows the quenched fractions in the TNG boxes for the $z=0$ population. This is defined as the fraction of quiescent galaxies of the entire (corresponding) population (based on a redshift-independent sSFR threshold $\log_{10}(\rm{sSFR}[y^{-1}])<-10.5$, see, e.g., \citealt{Lacerna2022}). Results are again qualitatively
consistent between boxes: a small fraction ($\lesssim 25\%$, depending on the box) of the population was already quenched at $z=1$. The quenched fraction increases towards $z=0$, reaching values between 60 and 75$\%$ (these results are in agreement with those from \citealt{Donnari2021a,Donnari2021b} using the IllustrisTNG simulation). Finally, in the lower panels of Fig. \ref{fig:sats_entire}, the description is complemented with the baryon-to-DM mass fraction. The relative amount of gas decreases steadily as the population consumes their gas reservoirs, from $M_{\rm gas}/M_{\rm DM} \sim 0.1$ at $z=1$ to $M_{\rm gas}/M_{\rm DM} \sim 0.02-0.05$ at $z=0$. The stellar-to-DM mass ratio consequently increases (almost by an order of magnitude, from $M_{\rm *}/M_{\rm DM} \sim 0.02-0.04$ to $\sim 0.1$).  

The fact that the TNG boxes map intrinsically different galaxy populations manifests itself in Fig. \ref{fig:sats_entire}. The larger the TNG box, the larger the mean stellar and subhalo masses of satellites in halos of a given mass. This selection effect implies that satellites appear to grow stellar mass more rapidly in the larger boxes, whereas the DM mass drops less abruptly. The size/resolution of the TNG box has also a clear impact on the fraction of quiescent galaxies: the quenched fraction is $\sim$25 $\%$ lower in the TNG300 box as compared to the TNG50 box. Note that one could expect that lower resolution makes the mass stripping process more efficient (i.e., larger ``chunks" of galaxies can be removed), but this is counteracted by the fact that TNG300 galaxies are more massive and (star-forming) active than TNG50 galaxies at fixed host halo mass. Interestingly, the evolution of the size ratio seems to be quite similar between boxes, independently of their distinct characteristics. In all boxes, $R_*/R_{\rm sh}$ increases by a factor of 3, resulting from the fact that the comoving stellar radius decreases less rapidly than the comoving subhalo radius. 

On average, the $z=0$ satellite population has grown mass at a decent rate within the redshift range considered (increasing by a factor 2-3 from $z=1$). It is important to stress that this selection does not impose any constraint on the past identity of $z=0$ satellites (some of them are centrals at previous times).
Results change significantly when we focus on a population that has suffered from mass stripping for a longer period of time (second panel of Fig. \ref{fig:sats_entire}, fiducial subset). Note that this population corresponds to 18, 21, and 25$\%$ of the entire $z=0$ population in the TNG50, TNG100, and TNG300 boxes, respectively. In TNG300, the average stellar mass raises slightly (up to $\sim 20 \%$ at $z \sim 0.5$), to then drop back to a value also slightly above the $z=1$ value at $z=0$. For TNG50/TNG100 populations, stellar mass actually decreases slightly within the redshift range considered. Note again that these average curves are the combined result of individual galaxies seeing their growth halted and even losing stellar mass at different times. 

The little evolution of stellar mass is accompanied by a more dramatic DM reduction than in the previous sample (by as much as $\sim 75\%$), and a slightly steeper increase in the size ratio, whose evolution is now dominated by the shrinking of the DM component. The effect of mass stripping is also significant for the gas-to-DM mass fraction, which drops by an order of magnitude, i.e., from 10 to 1$\%$. The $M_*/M_{\rm DM}$ is higher at $z=1$ than before (for the $z=0$ sample). As compared to the $z=0$ population, the quenched fraction of $z>1$ satellites is significantly higher for all boxes, indicating that mass stripping accelerates the quenching mechanism (at $z=0$, all $z>1$ satellites are in a quiescent state). Note that there are of course two main concurrent mechanisms that affect this evolution: mass stripping and star formation. Part of the gas depletion is due to galaxies turning gas into stars.

The connection between mass stripping and quenching becomes more clear when the evolution of quiescent and star-forming galaxies is addressed separately (from left to right, third and fourth plots of Fig. \ref{fig:sats_entire}, respectively). Here, we include satellites that remain quenched/star-forming during the entire redshift range (using the criteria explained above). Quenched galaxies lose stellar and DM mass steadily within the redshift range considered (up to $\sim 40$ and $\sim 70$ $\%$, respectively, at $z=0$). Note that {\it{we are subtracting the effect of mass growth due to star formation, so the effect of stellar mass stripping emerges clearly}}. Mass stripping in quenched galaxies produces variations in stellar mass, DM mass, and size ratios that are well represented by simple power laws of the form $\propto (1+z)^{\alpha}$. Finally, the lower panel shows that 
the stellar-to-DM mass ratio stays well above 0.1 throughout the redshift range considered for all boxes, increasing slightly due to the stronger DM stripping. As expected, the gas fraction is very low throughout. 

Star-forming galaxies, conversely, display a less uniform evolution. The average stellar mass increases steadily up to a factor of 3-4 within the redshift range considered, but DM mass remains almost constant until $z=0.4$. After this redshift, the DM fraction drops significantly, which result in a sudden increase of the stellar-to-subhalo size ratio. Here, subhaloes lose on average $\sim$ 50 to 65$\%$ of their DM content, but mass stripping is insufficient to produce any change in the stellar mass growth of the population. The baryon-to-DM mass fraction is as expected: high (slightly increasing) gas fraction due to the loss of DM and a strong increase of the stellar-to-DM fraction due to star formation and DM loss.

Another interesting conclusion from Fig. \ref{fig:sats_entire} is that, almost independently of the box chosen and the satellite galaxy selection, the evolution of the size ratio is quite similar (an increase of a  factor $\sim 3$ from $z=1$). In the following sections, we will dissect the effect of mass stripping in the $z>1$ satellite population. We will analyse the evolution of satellites as a function of both host halo mass and halo-centric distance. We will show that significant dependencies on the curves shown in Fig. \ref{fig:sats_entire} appear. Note that, statistically speaking, it is challenging to carry out this analysis controlling for stellar/subhalo mass (due to the mass ranges covered by the boxes). We have therefore opted to study the satellite population from TNG at ``face value" (i.e. avoiding attempting to study the exact same stellar/subhalo mass ranges), but the reader must keep in mind that some of the differences discussed are due to the varying selection across boxes.

\subsection{The evolution of satellite galaxies in lower-mass haloes}
\label{sec:mlow}

In order to study the effect of mass stripping in detail, we have determined the pericentric distance to the central galaxy ($d_{\rm peri}$) for each satellite in the fiducial sample. This is the {\it{minimum distance}} from the satellite to the corresponding central galaxy within the redshift range $0<z<1$ (it is, therefore, a single value within this interval). This and other similar measurements have proven more robust than the $z=0$ distance when it comes to evaluating the impact of the tidal field (e.g., satellites could be found far from the centre of the halo at $z=0$, despite having had a close encounter in the past).  
Four different subsets are defined based on normalised $d_{\rm peri}$ (pericentric distance divided by host halo virial radius): $d_{\rm peri} \le 0.1$, $0.1<d_{\rm peri} \le 0.4$, $0.4<d_{\rm peri} \le 0.75$, and $d_{\rm peri}>0.75$. In this section, we begin by analysing low-mass haloes in TNG. Fig. \ref{fig:sats_m1} presents, in the same format of Fig. \ref{fig:sats_entire}, the evolution of $z>1$ satellites in host haloes of masses $11 \le \log_{10}(M_{\rm host}[{\rm M_\odot}])<12$ for the four subsets described above. The subsets encompass in different boxes an average, from small to large $d_{\rm peri}$, of 13, 109, 46, and 24 objects, respectively. We also include the mean stellar mass of satellites in each subset. This is an important piece of information, since larger TNG boxes map typically more massive galaxies.

The effect of mass stripping is expected to become more noticeable as we move leftwards in Fig. \ref{fig:sats_m1}. Independently of the simulation box selected, the DM content decreases more rapidly for smaller $d_{\rm peri}$. For $d_{\rm peri}>0.75$, the DM fraction is reduced by half in TNG50, whereas no significant evolution is observed in TNG300. For $d_{\rm peri} \le 0.1$, conversely, DM is depleted to just a few percent of the $z=1$ mass content ($\sim$ 5$\%$ for TNG300 and $\lesssim$ 2$\%$ for TNG50). The progressive loss of DM for smaller distances results in a steeper evolution of the size ratio $R_{*}/R_{\rm sh}$.  

Statistics are modest in Fig. \ref{fig:sats_m1}, which explains the somewhat erratic behaviour of some of the curves shown. Although some differences exist, the growth in stellar mass appears fairly similar for different distances in all boxes. This evolution is very consistent with the quenched fractions. TNG300, which features the most massive satellites, displays the steeper stellar growth functions and, consequently, the lowest quenched fractions (barely any satellite is quenched at any redshift). The stellar mass growth flattens and the evolution of the quenched fraction steepens for progressively smaller TNG boxes. Fig. \ref{fig:sats_m1} therefore illustrates how differences between boxes at fixed halo mass and halo-centric distance can be attributed, at least partially, to systematically different galaxy/subhalo masses and quenched fractions. We will come back to this topic in Section \ref{sec:resolution}.

The cyan arrows in the top panel of Fig. \ref{fig:sats_m1} indicate the average pericentric redshift, $z_{\rm peri}$ (i.e., the average redshift at which galaxies have their pericentric passage), for the TNG50 (solid) and the TNG300 boxes (open). The pericentric passage, as defined in this work, typically occurs at recent times in this mass bin ($z_{\rm peri}<0.2$ for $d_{\rm peri}<0.75$). The $z_{\rm peri}$ redshift increases with $d_{\rm peri}$, which is consistent with the fact that objects with large $d_{\rm peri}$ take more time to reach smaller distances. Fig. \ref{fig:sats_m1} also shows that larger boxes have larger $z_{\rm peri}$. 

As mentioned before, Fig. \ref{fig:sats_m1} shows that DM tend to be stripped early on, which is consistent with the fact that it is stripped during the first close encounters 
\citep{Smith2016}. The magenta arrows indicate the average redshift at which the population reaches within a certain reference distance to the halo centre for the first time. This reference distance is set at 0.4$R_{\rm vir}$ in the first and second panels (from left to right) and 0.75$R_{\rm vir}$ in the third panel (no arrows are shown for the largest $d_{\rm peri}$). The magenta arrows indicate that satellites with the smallest $d_{\rm peri}$ also reach within close proximity to the central galaxy earlier on. The positions of the magenta arrows also roughly correlate with the redshifts at which DM starts to drop significantly, in agreement with the aforementioned results from \cite{Smith2016}. A visual inspection of the evolution of the halo-centric distance for individual objects (for different $d_{\rm peri}$ bins) reveals that objects with small pericentric distances typically have short-period spiralling orbits from $z=1$ or above, whereas large values of $d_{\rm peri}$ usually correspond to long-period orbits where the satellite is still far from merging.

To summarise our results so far, we find no conclusive signs of significant baryonic mass stripping in Fig. \ref{fig:sats_m1}, including gas and stellar mass. The evolution of these components seem to be dominated by star formation; only in TNG50, signs of an attenuated stellar mass growth start to be noticeable. Note that, even for $d_{\rm peri} \le 0.1$, where DM is reduced by more than a factor 10, the effect on stellar mass is almost unnoticeable. Connecting this result with the SsHMRs of Fig. \ref{fig:SHMR-evo}, it appears that the enhancement at the low-mass end could be at least partially driven by an excess of tidal stripping on the DM component of very small subhaloes (which has little effect on the stellar mass).

In Fig. \ref{fig:sats_m2}, we look into slightly more massive haloes, i.e., $12 \le \log_{10}(M_{\rm host}[{\rm M_\odot}])<13$, where the number of objects are on average 284, 1362, 454, and 154, for increasingly higher $d_{\rm peri}$ bins (the average is computed for the 3 boxes). The effect of mass stripping on the baryonic content of satellites here becomes much more evident, as a clear trend with distance (relative to halo size) emerges. For $d_{\rm peri}<0.1$, stellar mass grows up to a redshift that ranges from 0.7 for TNG50 to 0.3 to TNG300, only to subsequently decrease down to $z=0$. As we move to larger $d_{\rm peri}$, the effect of mass stripping hindering the growth of stellar mass is progressively reduced. For $d_{\rm peri}>0.75$, stellar mass seems to grow almost unaffectedly for all boxes. Again, $R_{*}/R_{\rm sh}$ increases more rapidly for smaller halo-centric distances, remaining quite consistent between boxes. Interestingly, despite the different evolution in stellar mass, the DM mass loss is quite similar to that reported for smaller haloes. 

\begin{figure*}
	\includegraphics[width=2.0\columnwidth]{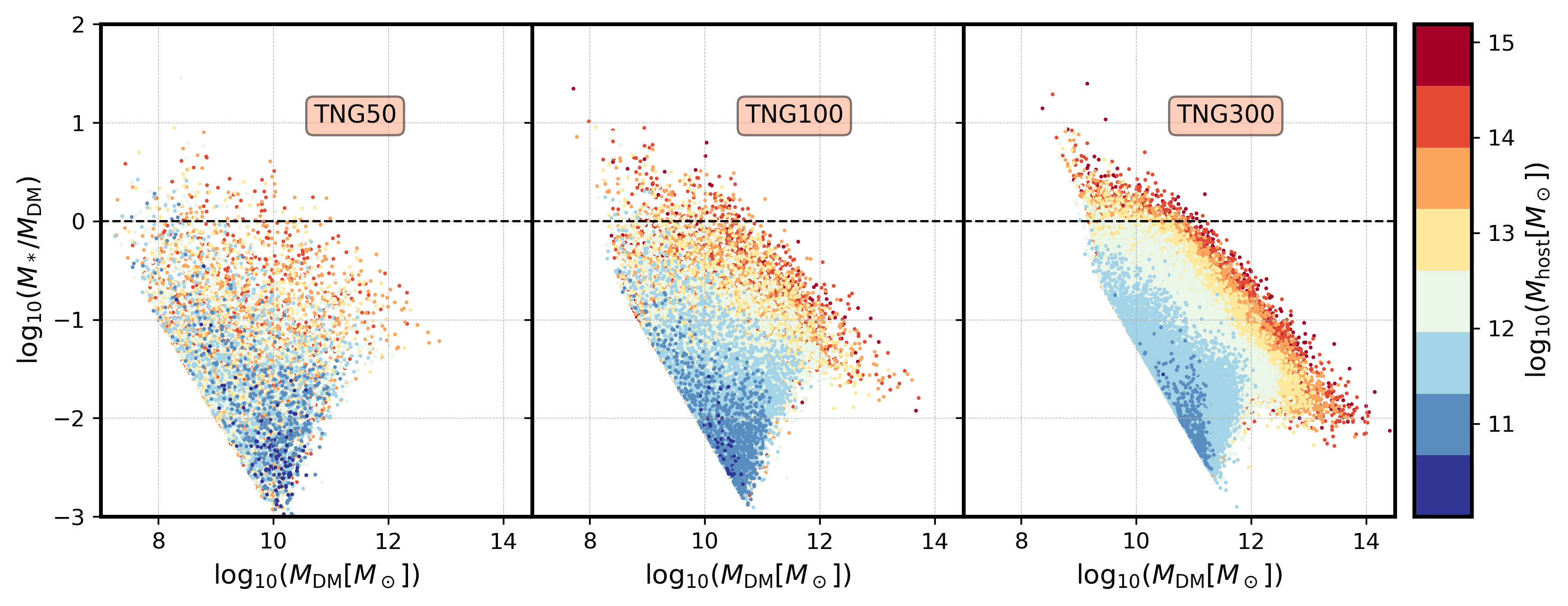}
    \caption{The stellar-to-DM mass ratio as a function of subhalo DM mass for all satellites at $z=0$ in TNG50, TNG100, and TNG300. The colour code indicates the mass of the host halo, as shown in the colour bar. The demarcation between normal and DM-poor subhaloes is represented by a dashed line.}
    \label{fig:noDMscatter}
\end{figure*}

\begin{figure}
    \begin{center}
	\includegraphics[width=0.7\columnwidth]{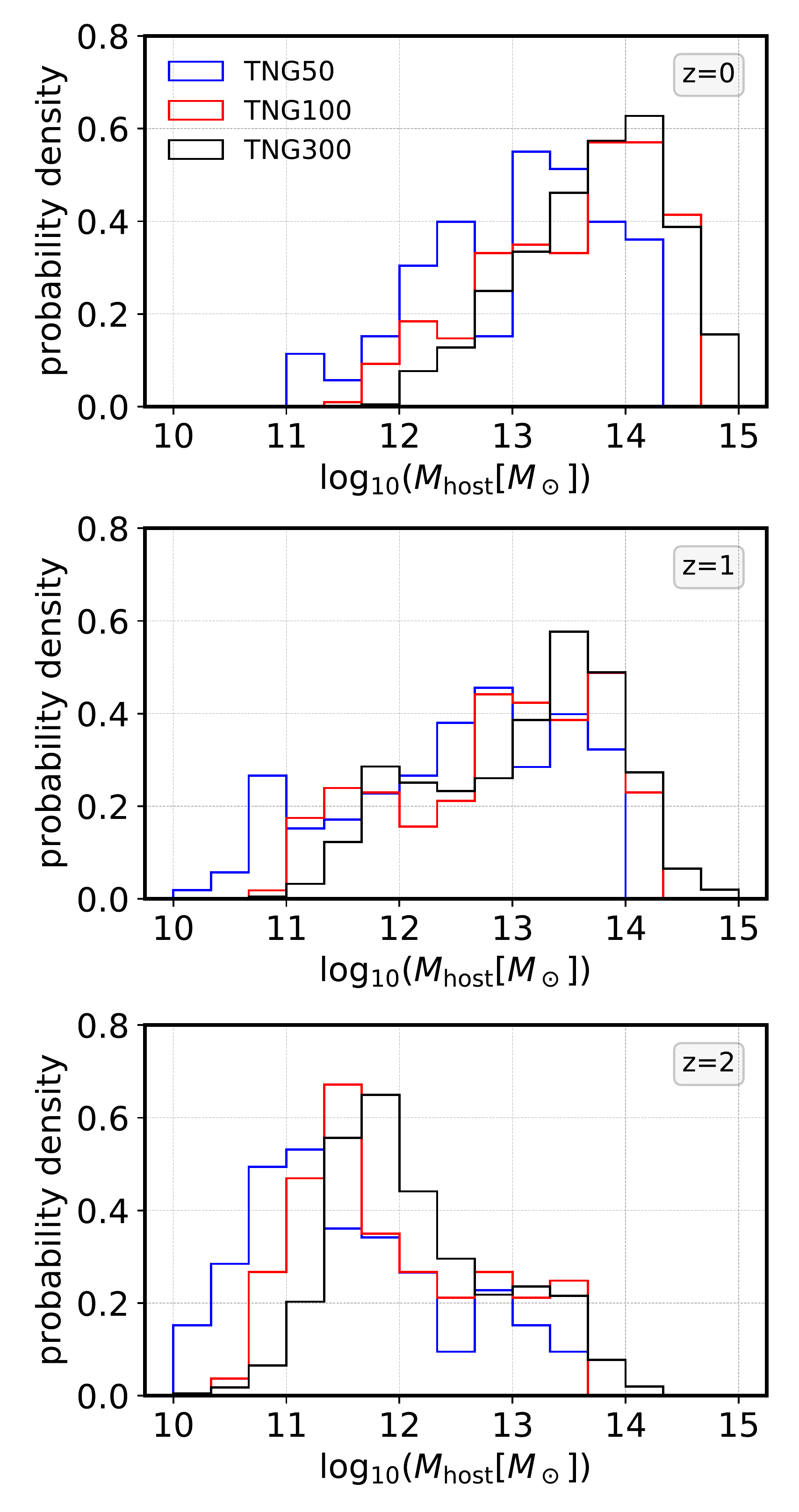}
    \caption{The distribution of haloes containing DM-deficient galaxies at $z=0$ in the TNG boxes (top) and their progenitors at $z=1$ (middle) and $z=2$ (bottom). Note that the progenitors of $z=0$ satellites are not necessarily satellites throughout the redshift range considered.}
   \label{fig:noDMhist}
   \end{center}
\end{figure}

\begin{figure}
	\includegraphics[width=1.0\columnwidth]{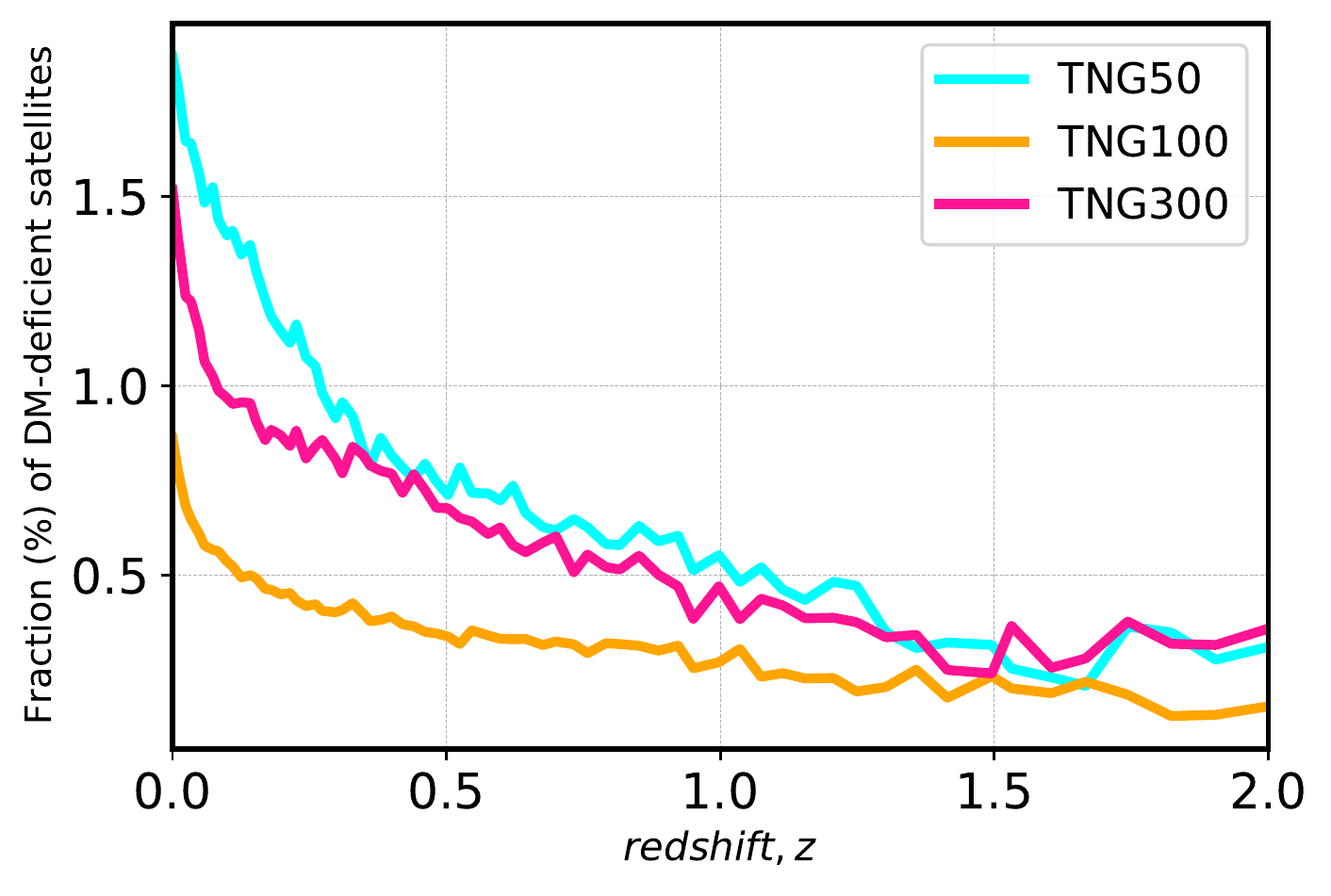}
    \caption{The redshift evolution of the fraction of DM-poor satellites (of the entire satellite population at each snapshot) for the three TNG boxes.}
    \label{fig:noDMfrac}
\end{figure}

An important result from Fig. \ref{fig:sats_m2} is the correlation between $d_{\rm peri}$, quenching, and stellar mass growth. For any given box, the closer the pericentric passage, the faster the satellite population quenches, and the more severe the mass growth attenuation. Interestingly, it is only when the quenched fraction reaches close to 100$\%$ that the average satellite ends up losing stellar mass (in a statistical sense). This (the stellar mass fraction being smaller than 1) only really happens for TNG50 and TNG100; in TNG300, the effect is not observed even when DM has dropped to less than 10$\%$ of its $z=1$ value. The effect of mass stripping is also noticeable from the lower panels of Fig. \ref{fig:sats_m2}. The quenching manifestation of mass stripping is of course connected to the gas mass fraction, which typically drops more rapidly as the population gets closer to the centre of the halo. This trend seems to be discontinued in the closest-distance bin, potentially due to resolution effects.  

The pericentric passage happens at slightly higher redshifts in Fig. \ref{fig:sats_m2}, as compared to Fig. \ref{fig:sats_m1}: $z_{\rm peri} \simeq 0.17$ for $d_{\rm peri}<0.1$ and $z_{\rm peri} \simeq 0.31$ for $d_{\rm peri}>0.75$ (on average for all boxes). The magenta arrows indicate that the first close encounter (defined through an arbitrary reference distance) happens at higher redshifts as well, as compared to Fig. \ref{fig:sats_m1}. This could provide an explanation as to why the effect of stellar mass stripping is noticeable in the leftmost panel of Fig. \ref{fig:sats_m2}, since satellites have been exposed to 
this mechanism for a longer period of time (as compared to satellites in Fig. \ref{fig:sats_m1}). Also note that we are studying the effects in terms of distances normalised to halo size. Haloes in Fig. \ref{fig:sats_m1} are typically physically smaller.

\subsection{The evolution of satellite galaxies in massive haloes}
\label{sec:mhigh}

Fig. \ref{fig:sats_m3} displays the evolution of satellite galaxies in host haloes of mass $13 \le \log_{10}(M_{\rm host}[{\rm M_\odot}])<14$, where the number of objects is on average 1379, 3544, 802, and 227, for increasingly higher $d_{\rm peri}$ bins. A good way to evaluate these results is to compare them with those for lower mass bins (Fig. \ref{fig:sats_m2}). Starting from the $d_{\rm peri}\le 0.1$ subset, we can see how the effect of mass stripping on the baryonic component of subhalos is more noticeable. For all boxes, the average stellar mass starts to decrease earlier and so does the gas-to-DM fraction. For host haloes of masses $12 \le \log_{10}(M_{\rm host}[{\rm M_\odot}])<13$, the gas fraction $M_{\rm gas}/M_{\rm DM}$ remains above $10^{-2}$, whereas for these more massive haloes it drops well below this threshold at $z=0$ ($M_{\rm gas}/M_{\rm DM} \simeq  3 \times 10^{-3} - 10^{-4}$, depending on the box). The quenching process is also significantly accelerated. 

Although the effect of quenching and stellar mass stripping is more noticeable, the loss of DM is, interestingly, less pronounced. It is therefore incorrect to assume that a more significant DM loss will also imply a flatter stellar mass growth, in the average population sense. Galaxies that experience the closest encounters in these haloes retain more that $10\%$ of their DM mass. We have checked that the average $d_{\rm peri}$ within the $d_{\rm peri}\le 0.1$ bin for these massive haloes is quite similar to that in less massive halos. Note, however, that distances are measured in virial radii, so a small $d_{\rm peri}$ for low mass haloes correspond to a smaller absolute distance. The fact that in smaller halos the orbits are more plunging (they reach closer to the central galaxy, in absolute terms) might explain the observed trends.

As a function of $d_{\rm peri}$, the trends that emerged in Fig. \ref{fig:sats_m2} remain in place, with the effect of mass stripping becoming less severe for higher distances: less quenching, higher gas fraction, more pronounced stellar mass growth, less DM loss. Importantly, the differences between boxes in each panel are significantly reduced as compared to lower halo mass bins, which implies that the effect of resolution and volume becomes less severe.  

In Fig. \ref{fig:sats_m4}, we look into the most massive haloes in the simulations: those in the halo mass range $14 \le \log_{10}(M_{\rm host}[{\rm M_\odot}])<15$ (1087, 2355, 425, 1012 objects on average between boxes in the $d_{\rm peri}$ subsets). Note that there are only a handful of subhaloes in the larger $d_{\rm peri}$ bins in TNG50, which explains the noisy results for this box. In order to avoid repetition, we will highlight two main conclusions from Fig. \ref{fig:sats_m4}. First, the overall trends with $d_{\rm peri}$ are still in place. Second, at fixed $d_{\rm peri}$, the loss of stellar and gas mass becomes even more significant, and, consequently, so does the quenching, as compared to lower mass haloes. DM loss is quite similar, although slightly less pronounced, particularly for the lowest $d_{\rm peri}$. 

Regarding the redshift of the pericentric and the first close encounters, the trends described in Section \ref{sec:mlow} still hold. At fixed $d_{\rm peri}$, satellites in more massive host haloes have their pericentric passage earlier on. In the most massive haloes (Fig. \ref{fig:sats_m4}), we find, on average: $z_{\rm peri} \simeq 0.27$ for $d_{\rm peri} \le 0.1$ and $z_{\rm peri} \simeq 0.42$ for $d_{\rm peri}>0.75$. On the other hand, satellites with $d_{\rm peri}\le 0.1$ in massive haloes were within 0.4$R_{\rm vir}$ already at $z \gtrsim 1$.

Finally, we have checked that, qualitatively, the main results of the paper remain valid when the $z=0$ halo-centric distance, instead of the pericentric distance, is employed. Interestingly, binning by $z=0$ distance produces slightly larger mass losses than binning by $d_{\rm peri}$ (both in stellar mass and DM mass). The reason behind this general agreement is that, as shown above, satellites tend to have spiralling orbits, so those objects with the smallest $d_{\rm peri}$ tend to reach proximity at recent times (also we have only analysed objects that remain satellites at $0<z<1$).

\subsection{Predictions controlling for stellar mass}
\label{sec:resolution}

Figs. \ref{fig:sats_m1} - \ref{fig:sats_m4} show that, qualitatively, the TNG boxes provide similar predictions for the average evolution of satellites, even when these are split by host halo mass and pericentric distance. Quantitatively, however, differences are apparent across the array of measurements performed in this analysis, due to the fact that boxes map different stellar mass (and subhalo mass) ranges. In this section we explicitly show this point: most of the differences found are the result of each box mapping intrinsically different populations in terms of stellar mass and/or subhalo DM mass.
 
In Fig. \ref{fig:tests}, we pick two different bins in host halo mass vs. distance space: $12\le\log_{10}(M_{\rm host}[{\rm M_\odot}])<13$, with $d_{\rm peri} \le 0.1$ (Fig. \ref{fig:sats_m2}) and  $13\le \log_{10}(M_{\rm host}[{\rm M_\odot}])<14$, with $0.1<d_{\rm peri} \le 0.4$ (Fig. \ref{fig:sats_m3}).
In both cases, we impose narrow $z=0$ stellar mass ranges of $9.2<\log_{10}(M_{\rm *}[{\rm M_\odot}])<9.5$ and $9<\log_{10}(M_{\rm *}[{\rm M_\odot}])<9.1$, respectively. This allows us to (approximately) fix the mean $z=0$ stellar mass in the three boxes, of course at the expense of losing statistics. Despite this caveat, Fig. \ref{fig:tests} shows that the differences observed in almost all properties are reduced considerably (particularly, as expected, for stellar mass growth and quenched fraction). In most cases, these differences are fairly consistent with the statistical error on the mean.

Note that there is of course very large intrinsic stochasticity in the evolution of galaxies, which is reflected in these properties. As an example, the standard deviations (object-to-object) in the $z=0$ stellar growth ratios are $\sim$0.46, 0.8, and 1.8 for the TNG50, TNG100, and TNG300 boxes, whereas the standard deviations for the $z=0$ DM fractions are $\sim$0.03, 0.1, and 0.06, respectively (even after controlling for stellar mass). The differences that still remain could also be due to the scatter in the SsHMR. 

Characterising the effect of varying resolution on the evolution of satellite galaxies as predicted by the TNG boxes is outside the scope of this paper, as it would require a more in depth analysis at the particle level. However, Fig. \ref{fig:tests} shows that differences between boxes are relatively small (given the intrinsic variation in the data) when the analysis is performed at fixed stellar mass. This allows us to use the boxes to evaluate the stellar mass dependence. Our results also suggest that the quenching would not depend strongly on resolution.

\begin{table*}
\begin{center}
\begin{tabular}{ccccccccc}        
\hline  Box &  $<\log_{10}(M_{\rm host})>$   &  $<\log_{10}(M_{\rm *})>$ & $<M_*/M_{\rm DM}>$ & $<R_*/R_{\rm sh}>$ & $\%$ of quenched & $<d_{\rm z=0}>$ & $<d_{\rm peri}>$ &  $<z_{\rm infall}>$\\
\hline
\hline  TNG50 & 13.097[0.825] & 9.076[0.83] & 1.985[1.24] & 0.884[0.128] & 88.0 & 0.36[0.602] & 0.054[0.055] & 1.438[0.576]\\
\hline  TNG100 & 13.535[0.772] & 9.660[0.637] & 2.087[1.782] & 0.891[0.107] & 87.7 & 0.437[0.936] & 0.048[0.066] & 1.443[0.592]\\
\hline  TNG300 & 13.763[0.671] & 9.946[0.412] &  1.591[1.279] & 0.915[0.11] & 94.1 & 0.342[0.825] & 0.035[0.044] & 1.339[0.581] \\
\hline  
\end{tabular}
\end{center}
\caption{Average values and standard deviations for several important properties of DM-deficient satellites at $z=0$ in the TNG boxes. From left to right: TNG box, mean host mass in solar masses, mean stellar mass in solar masses, mean stellar-to-DM mass ratio, mean stellar-to-subhalo size ratio, percentage of quenched galaxies, mean normalised halo-centric distance at $z=0$, mean normalised pericentric distance at $z=0$, and mean infall redshift.}
\label{table}  
\end{table*}

\subsection{An extreme case: the DM-deficient systems}
\label{sec:DM}

The effect of mass stripping is significantly stronger, as  Figs. \ref{fig:sats_m1} to \ref{fig:sats_m4} demonstrate, for the DM component of subhaloes than it is for stellar mass. This is expected, not only because the DM component is much more extended and less concentrated, but also because galaxies inside subhaloes continue forming stars (thus turning gas into stars). In this section, we evaluate the extreme case where subhaloes reach a point of DM loss where stellar mass becomes the dominant component. This analysis is connected with recent observational claims of the existence of galaxies containing no (or very little) DM (see \citealt{vD2018_noDM1,vD2019_noDM2}). 

The $z=0$ population of DM-deficient satellites is highlighted in the $M_*/M_{\rm DM }$ vs. $M_{\rm DM }$ plot of Fig. \ref{fig:noDMscatter}, as those objects lying above the horizontal dashed line. These galaxies are defined by the condition that the stellar-to-DM ratio is greater than 1. There are several important conclusions to extract from this figure. First, DM-deficient satellite galaxies are found in all TNG boxes. They correspond less than 1$\%$ of the TNG50, TNG100, and TNG300 galaxy populations at $z=0$. Second, they span a relatively wide range of subhalo DM masses, i.e., they are not only found in low-mass subhaloes, where resolution effects might be a concern. They, however, tend to avoid the most massive subhaloes, as expected from tidal stripping. Third, the most extreme cases, where the stellar-to-DM ratio reaches a factor as high as 10 tend to correspond to the smallest subhaloes. Finally, DM-deficient satellite galaxies also tend to reside in massive host haloes ($M_{\rm host} \gtrsim 10^{13}$ M$_\odot$). 

The latter aspect is addressed in more detail in Fig. \ref{fig:noDMhist}, where the probability density distribution of mass for haloes hosting the DM-deficient population at $z=0$ is displayed for the TNG boxes. We also show here the distribution of host haloes for the progenitors of the $z=0$ population, at redshifts $z=1$ and $z=2$. Fig. \ref{fig:noDMhist} shows that DM-poor galaxies tend to favour massive haloes even at $z=1$. Note that massive host haloes are less common, but they also contain more satellites. Another related question of interest is the location of these galaxies within their host haloes. We have estimated that the average halo-centric distance for DM-poor satellites at $z=0$ is $\sim$0.36, 0.44, and 0.34 virial radii in the TNG50, TNG100, and TNG300 boxes, respectively. However, most of these objects have experienced very close encounters in the past, as confirmed by their pericentric distances: $d_{\rm peri}\simeq$ 0.05, 0.05, and 0.03 $R_{200}$, respectively. They belong, therefore, to the closest distance bin defined in Figs. \ref{fig:sats_m1} to \ref{fig:sats_m4}.

We have also checked that all DM-deficient galaxies are indeed satellites at $z=0$ in the three TNG boxes. In fact, there are only a handful of instances along the merger trees of the three boxes where a central galaxy is DM-poor. In Fig. \ref{fig:noDMfrac}, the fraction of these systems of the total satellite population at each snapshot is displayed for the three boxes as a function of redshift. For all boxes, the fraction increases from $z=2$, reaching a discrete $\sim$1-2$\%$ at $z=0$. This evolution with redshift is totally consistent with the idea that these systems lose their DM through close encounters. These encounters are expected to be progressively more likely given the hierarchical nature of structure formation. 

Among central galaxies at $z=0$, the most DM deficient objects have $M_*/M_{\rm DM }$ ratios of 0.3-0.4. These objects are also special when it comes to their recent evolution. While more than 70$\%$ of centrals were already centrals at $z=2$, those objects for which $M_*/M_{\rm DM }>0.1$ only became centrals (for the last time) at (median and standard deviation) $z=0.11 \pm 0.17, 0.15 \pm 0.15$, and $0.08 \pm 0.06$ in the TNG50, TNG100, and TNG300 boxes, respectively. Again, the fact that they were recently satellites is consistent with the notion that DM deficiency is connected with tidal stripping. These objects can be associated with splashback galaxies (i.e., central galaxies at $z=0$ that live in the vicinity of massive haloes and were satellites in recent times).  

In Table \ref{table}, we list average values (and standard deviations) for some of the main properties of DM-deficient satellites in TNG. One interesting aspect to highlight is that the smaller TNG boxes (which have higher resolution) predict higher average stellar-to-DM mass fractions, reaching a factor of 2 for TNG50/TNG100 (this might be, however, affected by statistics, since these boxes have significantly fewer objects above the threshold). The size ratio, on the other hand, is  $R_*/R_{\rm sh} \sim 1$ in all boxes, showing how tidal stripping and interactions with the central galaxy has bared these subhaloes to the very core. The great majority of these galaxies are, as expected, quenched ($\sim 90 \%$). Finally, most of them have very high infall redshifts, around $z_{\rm infall} = 1.4$ (note that this is a lower limit, since we have only employed merger trees up to $z=2$). In fact, around $70\%$ of DM-poor satellites at $z=0$ in each box pertain to our fiducial subset of galaxies that were accreted at $z>1$ and remain satellites ever since. 

These results allow us to explain the presence of baryon-dominated galaxies as a result of the loss of material that happens in galaxy groups or clusters, without the need to invoke any mechanism that could potentially contradict the hierarchical merging scenario for galaxy formation. DM-deficient galaxies, such as the ones allegedly observed in \cite{vD2018_noDM1,vD2019_noDM2}, seem to have a pretty standard origin, representing simply the extreme tail of the tidally stripped satellite galaxy population.

It is important to stress that this analysis could be potentially affected {\it{from a quantitative standpoint}} by identification issues from the {\sc subfind} algorithm \citep{Muldrew2011,Springel2021}, which uses only 3D and not velocity information. 

\section{Summary and conclusions}
\label{sec:discussion}

As satellite galaxies orbit larger central galaxies, they suffer from a variety of physical processes that can alter their structure, mass content, and evolutionary path. In this work, we use the TNG hydrodynamical simulation to measure the average evolution of these objects from $z=1$ to $z=0$. The main goals of the paper are the following. First, we evaluate how mass stripping impacts the average evolution of satellites in different degrees depending on their stellar mass, the mass of the host halo, and the pericentric distance (the minimum distance to the central galaxy that a satellite/subhalo reaches within a certain period of time). Second, we compare the predictions from the three TNG boxes (TNG50, TNG100, TNG300), taking into account that they map significantly different stellar and subhalo mass ranges. Third, we identify and characterise DM-deficient galaxies, i.e., those galaxies containing more stellar than DM mass.

Before addressing the average evolution of individual satellites, we have focused on two of their main scaling relations. The halo mass -- galaxy size relation for TNG300 satellites at $z=0$ was presented in \cite{Rodriguez2021}, showing a very flat trend all the way up to $10^{15}\, {\rm M_\odot}$. As we extended this analysis to other boxes and redshifts, several conclusions are drawn. First, the halo mass -- galaxy size relation remains fairly flat for all redshifts in TNG300 and TNG100. TNG50 shows a rather erratic behaviour towards the low-mass end, particularly at low redshift. Second, volume and resolution have an undeniable impact on the measurement. In TNG, larger boxes typically contain larger subhaloes and galaxies, which dictates that the amplitude of the relation increases as a function of box size.  In terms of the redshift evolution, we again find some discrepancies: for TNG300, the amplitude of the relation decreases with redshift, a behaviour that is not so clearly observed in TNG50 and TNG100. 

We have also investigated the stellar-to-(sub)halo mass relation (SsHMR) of satellites, a measurement that has only recently been reported in the literature (see \citealt{Zhang2021, Engler2021}). We have shown a puzzling picture that emerges when the analysis is performed down to small subhalo masses. For subhaloes with masses above $\log_{10}(M_{\rm sh}[{\rm M_\odot}])\simeq 10$ for TNG50 and 11 for TNG300 (and somewhere in between for TNG100), the shape of the TNG SsHMR resembles that of central galaxies, with the peak of the stellar-to-subhalo mass fraction around $\log_{10}(M_{\rm sh}[{\rm M_\odot}])\simeq 12$. Some redshift evolution of the SsHMR is also found, particularly for the smaller boxes (the fraction tends to decrease with redshift). Below these characteristic masses, a large enhancement on the SsHMR is found for all boxes; this feature becomes less prominent as redshift increases. It is tempting to attribute this anomaly to DM mass stripping, since we subsequently show that the effect on DM is more severe for smaller subhaloes (whereas stellar mass tends to be preserved). However, the very low mass end of the SsHMR might still be affected by resolution, which forces us to be cautious on our interpretation (note that the effect is nevertheless 
observed at masses corresponding to a fairly large number of particles). Further investigation is needed to determine the nature of this feature. 

The main part of our analysis addresses the evolution of the properties of satellites from $z=1$, focusing on those that remain inside groups and clusters without losing their identity during this long period of time. These objects are therefore expected to be significantly affected by different mass stripping mechanisms. We have analysed 3 main factors that determine the severity of stripping: the mass of the host halo, the stellar/subhalo mass, and the normalised pericentric distance, $d_{\rm peri}$ (i.e., in units of virial radii). The main conclusions of this part of the analysis can be summarised as follows:

\begin{itemize}
    \item At fixed halo mass, the closer the pericentric passage relative to the virial radius of the host halo, the larger the fraction of DM lost by subaloes. Those that experience the closest encounters ($d_{\rm peri}\le 0.1\, R_{\rm vir}$) are left with anywhere between 20 to just a few percent of their $z=1$ DM mass, depending on halo mass. At fixed $d_{\rm peri}$  (relative to halo size), the DM component of satellites is less affected by mass stripping in massive host haloes than in their less massive counterparts, with the effect being particularly noticeable for very small $d_{\rm peri}$ values. 

    \item The stellar mass of satellites is also affected by mass stripping. This dictates that the stellar mass growth history of satellites is mostly the result of two counteracting processes: star formation and mass stripping. 
    
    \item Stellar mass stripping is less severe than DM mass stripping, which is in agreement with the picture that tidal stripping proceeds in an outside-in fashion (\citealt{Smith2016, Engler2021}). Below $\log_{10}(M_{\rm host}[{\rm M_\odot}])<12$, the stellar mass growth of the population is very mildly affected by stripping, independently of $d_{\rm peri}$. In more massive halos, the effect of mass stripping becomes noticeable, with the loss of stellar mass increasing with host halo mass and decreasing with $d_{\rm peri}$. In the most extreme case, for haloes of mass $14<\log_{10}(M_{\rm host}[{\rm M_\odot}])<15$, the average satellite with the closest encounters ($d_{\rm peri} \le 0.1$) loses 50$\%$ of its stellar mass by $z=0$.

    \item Satellites that have experienced closer pericentric passages also quench faster, at fixed halo mass. At fixed $d_{\rm peri}$, quenching increases with halo mass. As a reference, and independently of stellar mass, more than 70$\%$ of the satellites that experience the shortest $d_{\rm peri}$ in the most massive halos ($14<\log_{10}(M_{\rm host}[{\rm M_\odot}])<15$) are quenched by $z=0.4$.
    
    \item At fixed halo mass, satellites that reach smaller pericentric distances relative to halo size tend to have their pericentric passages later on. However, they get within a close distance to the centre of the halo earlier in time. By analysing individual orbits, we have checked that most of them correspond to spiralling trajectories where the satellite galaxy gets progressively closer to the central galaxy. 

    \item The stellar-to-subhalo size ratio ($R_{*}/R_{\rm sh}$) always increases with redshift. This evolution is steeper for smaller $d_{\rm peri}$ and smaller host haloes, as dictated by the effect of mass stripping on the DM component of the subhalo.

    \item All the above conclusions are qualitatively valid for the three TNG boxes. The quantitative differences that we find seem to originate from the stellar mass ranges mapped by different boxes. As a reference, the mean stellar mass for the entire $z>1$ population is $\log_{10}(M_{\rm *}[{\rm M_\odot}])=8.97$, 9.51, and 9.88 for TNG50, TNG100, and TNG300, respectively. 
    
    \item At fixed $d_{\rm peri}$ and host halo mass, for a given redshift, the more massive the galaxy population in TNG: 1) the less attenuated the stellar mass growths, 2) the less severe the DM mass stripping, 2) the smaller the fraction of quenched galaxies, 3) the larger the gas-to-DM mass ratio. The size ratio ($R_{*}/R_{\rm sh}$), on the other hand, depends on host halo mass: in massive haloes, it grows faster with time for more massive galaxies (i.e., in larger boxes), but the trend reverts for low-mass haloes.

\end{itemize}

We have also investigated the extreme case where subhaloes lose so much DM that stellar mass becomes a more dominant component. This analysis was motivated by recent observations suggesting this possibility (the NGC1052--DF2 and NGC1052--DF4 galaxies, see \citealt{vD2018_noDM1,vD2019_noDM2,vD2022}). We have shown that DM deficient systems appear in the three IllustrisTNG boxes and span a wide range of stellar and subhalo masses, confirming previous results using hydrodynamical simulations within the $\Lambda$CDM scenario \citep[see e.g.][]{Ploeckinger2018,Haslbauer2019,Haslbauer2019b,Sales2020,Applebaum2021}. These systems are, however, preferentially found inside fairly massive host haloes ($M_{\rm host} \gtrsim 10^{13}$ M$_\odot$). Importantly, they all have very close pericentric passages ($d_{\rm peri} \simeq 0.05$), reinforcing the idea that the tidal field of the host halo plays a fundamental role in their lack of DM. At $z=0$, they are found at a range of distances to the central galaxy (an average of $\sim 0.4\, R_{\rm vir }$ with large scatter). We have also shown that $z=0$ DM-deficient systems typically became satellites at high redshift,  $z_{\rm infall} \gtrsim 1.4$. Our work is complementary to that of \cite{Moreno2022}, who recently reported the discovery of a small set of satellites with similar properties to those of NGC1052--DF2 and NGC1052--DF4 in the high-resolution {\sc fire-2} simulation (similar resolution of TNG50). By using three simulation boxes, we have been able to expand this analysis to a wide range of masses and environments. 

In summary, our results provide a complete and very clear picture for the average evolution of satellites inside groups and clusters. This description provides fundamental information that can be incorporated in halo-galaxy connection models, which are key to the generation of galaxy mocks. In addition, our comparison between TNG boxes serve as a benchmark for future analyses of satellite galaxies using TNG and other hydrodynamical simulations. In terms of future work, the group finder of \cite{Rodriguez2020, Rodriguez2021} can be used to investigate the existence of DM-deficient systems, combining large spectroscopic surveys with information for fainter galaxies. The \citeauthor{Rodriguez2020} method, which includes an estimation of halo masses, can also be employed to measure the evolution of the scaling relations for satellites by combining spectroscopic surveys at different redshifts (an extension of \citealt{Rodriguez2021}).

\section*{Acknowledgements}

ADMD thanks Fondecyt for financial support through the Fondecyt Regular 2021 grant 1210612. FR thanks the support by Agencia Nacional de Promoci\'on Cient\'ifica y Tecno\'ologica, the Consejo Nacional de Investigaciones Cient\'{\i}ficas y T\'ecnicas (CONICET, Argentina) and the Secretar\'{\i}a de Ciencia y Tecnolog\'{\i}a de la Universidad Nacional de C\'ordoba (SeCyT-UNC, Argentina).
MCA acknowledges financial support from 
the  Seal of Excellence @UNIPD 2020 programme under the ACROGAL project. JCM acknowledges support from the European Union’s Horizon Europe research and innovation programme (COSMO-LYA, grant agreement 101044612). IFAE is partially funded by the CERCA program of the Generalitat de Catalunya. 
\section*{Data Availability}

The simulation data underlying this article are publicly available at the TNG website: https://www.tng-project.org/. The data results arising from this work will be shared on reasonable request to the corresponding authors. 



\bibliographystyle{mnras}
\bibliography{main} 

\begin{thebibliography}{}
\makeatletter
\relax
\def\mn@urlcharsother{\let\do\@makeother \do\$\do\&\do\#\do\^\do\_\do\%\do\~}
\def\mn@doi{\begingroup\mn@urlcharsother \@ifnextchar [ {\mn@doi@}
  {\mn@doi@[]}}
\def\mn@doi@[#1]#2{\def\@tempa{#1}\ifx\@tempa\@empty \href
  {http://dx.doi.org/#2} {doi:#2}\else \href {http://dx.doi.org/#2} {#1}\fi
  \endgroup}
\def\mn@eprint#1#2{\mn@eprint@#1:#2::\@nil}
\def\mn@eprint@arXiv#1{\href {http://arxiv.org/abs/#1} {{\tt arXiv:#1}}}
\def\mn@eprint@dblp#1{\href {http://dblp.uni-trier.de/rec/bibtex/#1.xml}
  {dblp:#1}}
\def\mn@eprint@#1:#2:#3:#4\@nil{\def\@tempa {#1}\def\@tempb {#2}\def\@tempc
  {#3}\ifx \@tempc \@empty \let \@tempc \@tempb \let \@tempb \@tempa \fi \ifx
  \@tempb \@empty \def\@tempb {arXiv}\fi \@ifundefined
  {mn@eprint@\@tempb}{\@tempb:\@tempc}{\expandafter \expandafter \csname
  mn@eprint@\@tempb\endcsname \expandafter{\@tempc}}}

\bibitem[\protect\citeauthoryear{Abadi, Moore  \& Bower}{Abadi
  et~al.}{1999}]{Abadi1999}
Abadi M.~G.,  Moore B.,   Bower R.~G.,  1999, Monthly Notices of the Royal
  Astronomical Society, 308, 947

\bibitem[\protect\citeauthoryear{Abazajian et~al.,}{Abazajian
  et~al.}{2009}]{Abazajian2009}
Abazajian K.~N.,  et~al., 2009, The Astrophysical Journal Supplement Series,
  182, 543

\bibitem[\protect\citeauthoryear{{Applebaum}, {Brooks}, {Christensen},
  {Munshi}, {Quinn}, {Shen}  \& {Tremmel}}{{Applebaum}
  et~al.}{2021}]{Applebaum2021}
{Applebaum} E.,  {Brooks} A.~M.,  {Christensen} C.~R.,  {Munshi} F.,  {Quinn}
  T.~R.,  {Shen} S.,   {Tremmel} M.,  2021, \mn@doi [\apj]
  {10.3847/1538-4357/abcafa}, \href
  {https://ui.adsabs.harvard.edu/abs/2021ApJ...906...96A} {906, 96}

\bibitem[\protect\citeauthoryear{Balogh, Navarro  \& Morris}{Balogh
  et~al.}{2000}]{Balogh2000}
Balogh M.~L.,  Navarro J.~F.,   Morris S.~L.,  2000, The Astrophysical Journal,
  540, 113

\bibitem[\protect\citeauthoryear{Behroozi, Conroy  \& Wechsler}{Behroozi
  et~al.}{2010}]{Behroozi2010}
Behroozi P.~S.,  Conroy C.,   Wechsler R.~H.,  2010, The Astrophysical Journal,
  717, 379

\bibitem[\protect\citeauthoryear{{Beltz-Mohrmann}, {Berlind}  \&
  {Szewciw}}{{Beltz-Mohrmann} et~al.}{2020}]{Beltz-Mohrmann2020}
{Beltz-Mohrmann} G.~D.,  {Berlind} A.~A.,   {Szewciw} A.~O.,  2020, \mn@doi
  [\mnras] {10.1093/mnras/stz3442}, \href
  {https://ui.adsabs.harvard.edu/abs/2020MNRAS.491.5771B} {491, 5771}

\bibitem[\protect\citeauthoryear{{Bose}, {Eisenstein}, {Hernquist},
  {Pillepich}, {Nelson}, {Marinacci}, {Springel}  \& {Vogelsberger}}{{Bose}
  et~al.}{2019}]{Bose2019}
{Bose} S.,  {Eisenstein} D.~J.,  {Hernquist} L.,  {Pillepich} A.,  {Nelson} D.,
   {Marinacci} F.,  {Springel} V.,   {Vogelsberger} M.,  2019, \mn@doi [\mnras]
  {10.1093/mnras/stz2546}, \href
  {https://ui.adsabs.harvard.edu/abs/2019MNRAS.tmp.2192B} {p.~2192}

\bibitem[\protect\citeauthoryear{{Boselli} et~al.,}{{Boselli}
  et~al.}{2014}]{Boselli2014}
{Boselli} A.,  et~al., 2014, \mn@doi [\aap] {10.1051/0004-6361/201424419},
  \href {https://ui.adsabs.harvard.edu/abs/2014A&A...570A..69B} {570, A69}

\bibitem[\protect\citeauthoryear{{Boselli}, {Fossati}  \& {Sun}}{{Boselli}
  et~al.}{2022}]{Boselli2022}
{Boselli} A.,  {Fossati} M.,   {Sun} M.,  2022, \mn@doi [\aapr]
  {10.1007/s00159-022-00140-3}, \href
  {https://ui.adsabs.harvard.edu/abs/2022A&ARv..30....3B} {30, 3}

\bibitem[\protect\citeauthoryear{{Boylan-Kolchin}, {Ma}  \&
  {Quataert}}{{Boylan-Kolchin} et~al.}{2008}]{BoylanKolchin2008}
{Boylan-Kolchin} M.,  {Ma} C.-P.,   {Quataert} E.,  2008, \mn@doi [\mnras]
  {10.1111/j.1365-2966.2007.12530.x}, \href
  {https://ui.adsabs.harvard.edu/abs/2008MNRAS.383...93B} {383, 93}

\bibitem[\protect\citeauthoryear{Campbell, van~den Bosch, Hearin, Padmanabhan,
  Berlind, Mo, Tinker  \& Yang}{Campbell et~al.}{2015}]{Campbell2015}
Campbell D.,  van~den Bosch F.~C.,  Hearin A.,  Padmanabhan N.,  Berlind A.,
  Mo H.,  Tinker J.,   Yang X.,  2015, Monthly Notices of the Royal
  Astronomical Society, 452, 444

\bibitem[\protect\citeauthoryear{{Chandrasekhar}}{{Chandrasekhar}}{1943}]{Chandrasekhar1943}
{Chandrasekhar} S.,  1943, \mn@doi [\apj] {10.1086/144517}, \href
  {https://ui.adsabs.harvard.edu/abs/1943ApJ....97..255C} {97, 255}

\bibitem[\protect\citeauthoryear{{Chaves-Montero}, {Angulo}, {Schaye},
  {Schaller}, {Crain}, {Furlong}  \& {Theuns}}{{Chaves-Montero}
  et~al.}{2016}]{ChavesMontero2016}
{Chaves-Montero} J.,  {Angulo} R.~E.,  {Schaye} J.,  {Schaller} M.,  {Crain}
  R.~A.,  {Furlong} M.,   {Theuns} T.,  2016, \mn@doi [\mnras]
  {10.1093/mnras/stw1225}, \href
  {https://ui.adsabs.harvard.edu/abs/2016MNRAS.460.3100C} {460, 3100}

\bibitem[\protect\citeauthoryear{{Contreras}, {Angulo}  \&
  {Zennaro}}{{Contreras} et~al.}{2020}]{Contreras2020}
{Contreras} S.,  {Angulo} R.,   {Zennaro} M.,  2020, arXiv e-prints, \href
  {https://ui.adsabs.harvard.edu/abs/2020arXiv200503672C} {p. arXiv:2005.03672}

\bibitem[\protect\citeauthoryear{{Cybulski}, {Yun}, {Fazio}  \&
  {Gutermuth}}{{Cybulski} et~al.}{2014}]{Cybulski2014}
{Cybulski} R.,  {Yun} M.~S.,  {Fazio} G.~G.,   {Gutermuth} R.~A.,  2014,
  \mn@doi [\mnras] {10.1093/mnras/stu200}, \href
  {https://ui.adsabs.harvard.edu/abs/2014MNRAS.439.3564C} {439, 3564}

\bibitem[\protect\citeauthoryear{{Davis}, {Efstathiou}, {Frenk}  \&
  {White}}{{Davis} et~al.}{1985}]{Davis1985}
{Davis} M.,  {Efstathiou} G.,  {Frenk} C.~S.,   {White} S.~D.~M.,  1985,
  \mn@doi [\apj] {10.1086/163168}, \href
  {http://adsabs.harvard.edu/abs/1985ApJ...292..371D} {292, 371}

\bibitem[\protect\citeauthoryear{{Donnari} et~al.,}{{Donnari}
  et~al.}{2021a}]{Donnari2021a}
{Donnari} M.,  et~al., 2021a, \mn@doi [\mnras] {10.1093/mnras/staa3006}, \href
  {https://ui.adsabs.harvard.edu/abs/2021MNRAS.500.4004D} {500, 4004}

\bibitem[\protect\citeauthoryear{{Donnari}, {Pillepich}, {Nelson}, {Marinacci},
  {Vogelsberger}  \& {Hernquist}}{{Donnari} et~al.}{2021b}]{Donnari2021b}
{Donnari} M.,  {Pillepich} A.,  {Nelson} D.,  {Marinacci} F.,  {Vogelsberger}
  M.,   {Hernquist} L.,  2021b, \mn@doi [\mnras] {10.1093/mnras/stab1950},
  \href {https://ui.adsabs.harvard.edu/abs/2021MNRAS.506.4760D} {506, 4760}

\bibitem[\protect\citeauthoryear{{Dressler}}{{Dressler}}{1980}]{Dressler1980}
{Dressler} A.,  1980, \mn@doi [\apj] {10.1086/157753}, \href
  {https://ui.adsabs.harvard.edu/abs/1980ApJ...236..351D} {236, 351}

\bibitem[\protect\citeauthoryear{{Engler} et~al.,}{{Engler}
  et~al.}{2021}]{Engler2021}
{Engler} C.,  et~al., 2021, \mn@doi [\mnras] {10.1093/mnras/staa3505}, \href
  {https://ui.adsabs.harvard.edu/abs/2021MNRAS.500.3957E} {500, 3957}

\bibitem[\protect\citeauthoryear{{Favole}, {Montero-Dorta}, {Artale},
  {Contreras}, {Zehavi}  \& {Xu}}{{Favole} et~al.}{2022}]{Favole2022}
{Favole} G.,  {Montero-Dorta} A.~D.,  {Artale} M.~C.,  {Contreras} S.,
  {Zehavi} I.,   {Xu} X.,  2022, \mn@doi [\mnras] {10.1093/mnras/stab3006},
  \href {https://ui.adsabs.harvard.edu/abs/2022MNRAS.509.1614F} {509, 1614}

\bibitem[\protect\citeauthoryear{{Fillingham}, {Cooper}, {Pace},
  {Boylan-Kolchin}, {Bullock}, {Garrison-Kimmel}  \& {Wheeler}}{{Fillingham}
  et~al.}{2016}]{Fillingham2016}
{Fillingham} S.~P.,  {Cooper} M.~C.,  {Pace} A.~B.,  {Boylan-Kolchin} M.,
  {Bullock} J.~S.,  {Garrison-Kimmel} S.,   {Wheeler} C.,  2016, \mn@doi
  [\mnras] {10.1093/mnras/stw2131}, \href
  {https://ui.adsabs.harvard.edu/abs/2016MNRAS.463.1916F} {463, 1916}

\bibitem[\protect\citeauthoryear{{Fillingham}, {Cooper}, {Boylan-Kolchin},
  {Bullock}, {Garrison-Kimmel}  \& {Wheeler}}{{Fillingham}
  et~al.}{2018}]{Fillingham2018}
{Fillingham} S.~P.,  {Cooper} M.~C.,  {Boylan-Kolchin} M.,  {Bullock} J.~S.,
  {Garrison-Kimmel} S.,   {Wheeler} C.,  2018, \mn@doi [\mnras]
  {10.1093/mnras/sty958}, \href
  {https://ui.adsabs.harvard.edu/abs/2018MNRAS.477.4491F} {477, 4491}

\bibitem[\protect\citeauthoryear{{Genel} et~al.,}{{Genel}
  et~al.}{2018}]{Genel2018}
{Genel} S.,  et~al., 2018, \mn@doi [\mnras] {10.1093/mnras/stx3078}, \href
  {https://ui.adsabs.harvard.edu/abs/2018MNRAS.474.3976G} {474, 3976}

\bibitem[\protect\citeauthoryear{{Gnedin}, {Kravtsov}, {Klypin}  \&
  {Nagai}}{{Gnedin} et~al.}{2004}]{Gnedin2004}
{Gnedin} O.~Y.,  {Kravtsov} A.~V.,  {Klypin} A.~A.,   {Nagai} D.,  2004,
  \mn@doi [\apj] {10.1086/424914}, \href
  {https://ui.adsabs.harvard.edu/abs/2004ApJ...616...16G} {616, 16}

\bibitem[\protect\citeauthoryear{{Gu} et~al.,}{{Gu} et~al.}{2020}]{Gu2020}
{Gu} M.,  et~al., 2020, arXiv e-prints, \href
  {https://ui.adsabs.harvard.edu/abs/2020arXiv201004166G} {p. arXiv:2010.04166}

\bibitem[\protect\citeauthoryear{Gunn \& Gott}{Gunn \& Gott}{1972}]{Gunn1972}
Gunn J.~E.,  Gott J.~R.,  1972, The Astrophysical Journal, 176, 1

\bibitem[\protect\citeauthoryear{{Hadzhiyska}, {Bose}, {Eisenstein},
  {Hernquist}  \& {Spergel}}{{Hadzhiyska} et~al.}{2020}]{Hadzhiyska2020}
{Hadzhiyska} B.,  {Bose} S.,  {Eisenstein} D.,  {Hernquist} L.,   {Spergel}
  D.~N.,  2020, \mn@doi [\mnras] {10.1093/mnras/staa623}, \href
  {https://ui.adsabs.harvard.edu/abs/2020MNRAS.493.5506H} {493, 5506}

\bibitem[\protect\citeauthoryear{{Hadzhiyska}, {Bose}, {Eisenstein}  \&
  {Hernquist}}{{Hadzhiyska} et~al.}{2021}]{Hadzhiyska2021}
{Hadzhiyska} B.,  {Bose} S.,  {Eisenstein} D.,   {Hernquist} L.,  2021, \mn@doi
  [\mnras] {10.1093/mnras/staa3776}, \href
  {https://ui.adsabs.harvard.edu/abs/2021MNRAS.501.1603H} {501, 1603}

\bibitem[\protect\citeauthoryear{{Haslbauer}, {Banik}, {Kroupa}  \&
  {Grishunin}}{{Haslbauer} et~al.}{2019a}]{Haslbauer2019}
{Haslbauer} M.,  {Banik} I.,  {Kroupa} P.,   {Grishunin} K.,  2019a, \mn@doi
  [\mnras] {10.1093/mnras/stz2270}, \href
  {https://ui.adsabs.harvard.edu/abs/2019MNRAS.489.2634H} {489, 2634}

\bibitem[\protect\citeauthoryear{{Haslbauer}, {Dabringhausen}, {Kroupa},
  {Javanmardi}  \& {Banik}}{{Haslbauer} et~al.}{2019b}]{Haslbauer2019b}
{Haslbauer} M.,  {Dabringhausen} J.,  {Kroupa} P.,  {Javanmardi} B.,   {Banik}
  I.,  2019b, \mn@doi [\aap] {10.1051/0004-6361/201833771}, \href
  {https://ui.adsabs.harvard.edu/abs/2019A&A...626A..47H} {626, A47}

\bibitem[\protect\citeauthoryear{Hearin, Behroozi, Kravtsov  \& Moster}{Hearin
  et~al.}{2019}]{Hearin2019}
Hearin A.,  Behroozi P.,  Kravtsov A.,   Moster B.,  2019, Monthly Notices of
  the Royal Astronomical Society, 489, 1805

\bibitem[\protect\citeauthoryear{Jiang, Dekel, Freundlich, Bosch, Green,
  Hopkins, Benson  \& Du}{Jiang et~al.}{2020}]{Jiang2020}
Jiang F.,  Dekel A.,  Freundlich J.,  Bosch F.~C.,  Green S.~B.,  Hopkins
  P.~F.,  Benson A.,   Du X.,  2020, arXiv preprint arXiv:2005.05974

\bibitem[\protect\citeauthoryear{Kravtsov}{Kravtsov}{2013}]{Kravtsov2013}
Kravtsov A.~V.,  2013, The Astrophysical Journal Letters, 764, L31

\bibitem[\protect\citeauthoryear{{Lacerna} et~al.,}{{Lacerna}
  et~al.}{2022}]{Lacerna2022}
{Lacerna} I.,  et~al., 2022, \mn@doi [\mnras] {10.1093/mnras/stac1020}, \href
  {https://ui.adsabs.harvard.edu/abs/2022MNRAS.513.2271L} {513, 2271}

\bibitem[\protect\citeauthoryear{Larson, Tinsley  \& Caldwell}{Larson
  et~al.}{1980}]{Larson1980}
Larson R.,  Tinsley B.,   Caldwell C.~N.,  1980, The Astrophysical Journal,
  237, 692

\bibitem[\protect\citeauthoryear{{Mandelbaum}, {Seljak}, {Kauffmann}, {Hirata}
  \& {Brinkmann}}{{Mandelbaum} et~al.}{2006}]{Mandelbaum2006}
{Mandelbaum} R.,  {Seljak} U.,  {Kauffmann} G.,  {Hirata} C.~M.,   {Brinkmann}
  J.,  2006, \mn@doi [\mnras] {10.1111/j.1365-2966.2006.10156.x}, \href
  {https://ui.adsabs.harvard.edu/abs/2006MNRAS.368..715M} {368, 715}

\bibitem[\protect\citeauthoryear{{Marinacci} et~al.,}{{Marinacci}
  et~al.}{2018}]{Marinacci2018}
{Marinacci} F.,  et~al., 2018, \mn@doi [\mnras] {10.1093/mnras/sty2206}, \href
  {https://ui.adsabs.harvard.edu/abs/2018MNRAS.480.5113M} {480, 5113}

\bibitem[\protect\citeauthoryear{Mastropietro, Moore, Mayer, Debattista,
  Piffaretti  \& Stadel}{Mastropietro et~al.}{2005}]{Mastropietro2005}
Mastropietro C.,  Moore B.,  Mayer L.,  Debattista V.~P.,  Piffaretti R.,
  Stadel J.,  2005, Monthly Notices of the Royal Astronomical Society, 364, 607

\bibitem[\protect\citeauthoryear{{Mayer}, {Mastropietro}, {Wadsley}, {Stadel}
  \& {Moore}}{{Mayer} et~al.}{2006}]{Meyer2006}
{Mayer} L.,  {Mastropietro} C.,  {Wadsley} J.,  {Stadel} J.,   {Moore} B.,
  2006, \mn@doi [\mnras] {10.1111/j.1365-2966.2006.10403.x}, \href
  {https://ui.adsabs.harvard.edu/abs/2006MNRAS.369.1021M} {369, 1021}

\bibitem[\protect\citeauthoryear{McCarthy, Frenk, Font, Lacey, Bower, Mitchell,
  Balogh  \& Theuns}{McCarthy et~al.}{2008}]{Mccarthy2008}
McCarthy I.~G.,  Frenk C.~S.,  Font A.~S.,  Lacey C.~G.,  Bower R.~G.,
  Mitchell N.~L.,  Balogh M.~L.,   Theuns T.,  2008, Monthly Notices of the
  Royal Astronomical Society, 383, 593

\bibitem[\protect\citeauthoryear{{Merritt}}{{Merritt}}{1983}]{Merritt1983}
{Merritt} D.,  1983, \mn@doi [\apj] {10.1086/160571}, \href
  {https://ui.adsabs.harvard.edu/abs/1983ApJ...264...24M} {264, 24}

\bibitem[\protect\citeauthoryear{Merritt}{Merritt}{1984}]{Merritt1984}
Merritt D.,  1984, The Astrophysical Journal, 276, 26

\bibitem[\protect\citeauthoryear{{Merritt}}{{Merritt}}{1985}]{Merritt1985}
{Merritt} D.,  1985, \mn@doi [\apj] {10.1086/162860}, \href
  {https://ui.adsabs.harvard.edu/abs/1985ApJ...289...18M} {289, 18}

\bibitem[\protect\citeauthoryear{{Montero-Dorta} et~al.,}{{Montero-Dorta}
  et~al.}{2020}]{MonteroDorta2020A}
{Montero-Dorta} A.~D.,  et~al., 2020, \mn@doi [\mnras]
  {10.1093/mnras/staa1624}, \href
  {https://ui.adsabs.harvard.edu/abs/2020MNRAS.496.1182M} {496, 1182}

\bibitem[\protect\citeauthoryear{{Montero-Dorta}, {Artale}, {Abramo}  \&
  {Tucci}}{{Montero-Dorta} et~al.}{2021a}]{MonteroDorta2021A}
{Montero-Dorta} A.~D.,  {Artale} M.~C.,  {Abramo} L.~R.,   {Tucci} B.,  2021a,
  \mn@doi [\mnras] {10.1093/mnras/stab1026}, \href
  {https://ui.adsabs.harvard.edu/abs/2021MNRAS.504.4568M} {504, 4568}

\bibitem[\protect\citeauthoryear{{Montero-Dorta}, {Chaves-Montero}, {Artale}
  \& {Favole}}{{Montero-Dorta} et~al.}{2021b}]{MonteroDorta2021C}
{Montero-Dorta} A.~D.,  {Chaves-Montero} J.,  {Artale} M.~C.,   {Favole} G.,
  2021b, \mn@doi [\mnras] {10.1093/mnras/stab2556}, \href
  {https://ui.adsabs.harvard.edu/abs/2021MNRAS.508..940M} {508, 940}

\bibitem[\protect\citeauthoryear{{Moore}, {Katz}, {Lake}, {Dressler}  \&
  {Oemler}}{{Moore} et~al.}{1996}]{Moore1996}
{Moore} B.,  {Katz} N.,  {Lake} G.,  {Dressler} A.,   {Oemler} A.,  1996,
  \mn@doi [\nat] {10.1038/379613a0}, \href
  {https://ui.adsabs.harvard.edu/abs/1996Natur.379..613M} {379, 613}

\bibitem[\protect\citeauthoryear{{Moore}, {Lake}  \& {Katz}}{{Moore}
  et~al.}{1998}]{Moore1998}
{Moore} B.,  {Lake} G.,   {Katz} N.,  1998, \mn@doi [\apj] {10.1086/305264},
  \href {https://ui.adsabs.harvard.edu/abs/1998ApJ...495..139M} {495, 139}

\bibitem[\protect\citeauthoryear{Moore, Lake, Quinn  \& Stadel}{Moore
  et~al.}{1999}]{Moore1999}
Moore B.,  Lake G.,  Quinn T.,   Stadel J.,  1999, Monthly Notices of the Royal
  Astronomical Society, 304, 465

\bibitem[\protect\citeauthoryear{{Moreno} et~al.,}{{Moreno}
  et~al.}{2022}]{Moreno2022}
{Moreno} J.,  et~al., 2022, \mn@doi [Nature Astronomy]
  {10.1038/s41550-021-01598-4}, \href
  {https://ui.adsabs.harvard.edu/abs/2022NatAs...6..496M} {6, 496}

\bibitem[\protect\citeauthoryear{Moster, Somerville, Maulbetsch, van~den Bosch,
  Macci{\`{o}}, Naab  \& Oser}{Moster et~al.}{2010}]{Moster2010}
Moster B.~P.,  Somerville R.~S.,  Maulbetsch C.,  van~den Bosch F.~C.,
  Macci{\`{o}} A.~V.,  Naab T.,   Oser L.,  2010, \mn@doi [The Astrophysical
  Journal] {10.1088/0004-637x/710/2/903}, 710, 903

\bibitem[\protect\citeauthoryear{{Muldrew}, {Pearce}  \& {Power}}{{Muldrew}
  et~al.}{2011}]{Muldrew2011}
{Muldrew} S.~I.,  {Pearce} F.~R.,   {Power} C.,  2011, \mn@doi [\mnras]
  {10.1111/j.1365-2966.2010.17636.x}, \href
  {https://ui.adsabs.harvard.edu/abs/2011MNRAS.410.2617M} {410, 2617}

\bibitem[\protect\citeauthoryear{{Naab} \& {Ostriker}}{{Naab} \&
  {Ostriker}}{2017}]{Naab2017}
{Naab} T.,  {Ostriker} J.~P.,  2017, \mn@doi [\araa]
  {10.1146/annurev-astro-081913-040019}, \href
  {https://ui.adsabs.harvard.edu/abs/2017ARA&A..55...59N} {55, 59}

\bibitem[\protect\citeauthoryear{{Nelson} et~al.,}{{Nelson}
  et~al.}{2018}]{Nelson2018_ColorBim}
{Nelson} D.,  et~al., 2018, \mn@doi [\mnras] {10.1093/mnras/stx3040}, \href
  {https://ui.adsabs.harvard.edu/abs/2018MNRAS.475..624N} {475, 624}

\bibitem[\protect\citeauthoryear{{Nelson} et~al.,}{{Nelson}
  et~al.}{2019}]{Nelson2019}
{Nelson} D.,  et~al., 2019, \mn@doi [Computational Astrophysics and Cosmology]
  {10.1186/s40668-019-0028-x}, \href
  {https://ui.adsabs.harvard.edu/abs/2019ComAC...6....2N} {6, 2}

\bibitem[\protect\citeauthoryear{{Pillepich} et~al.,}{{Pillepich}
  et~al.}{2018a}]{Pillepich2018}
{Pillepich} A.,  et~al., 2018a, \mn@doi [\mnras] {10.1093/mnras/stx2656}, \href
  {https://ui.adsabs.harvard.edu/abs/2018MNRAS.473.4077P} {473, 4077}

\bibitem[\protect\citeauthoryear{{Pillepich} et~al.,}{{Pillepich}
  et~al.}{2018b}]{Pillepich2018b}
{Pillepich} A.,  et~al., 2018b, \mn@doi [\mnras] {10.1093/mnras/stx3112}, \href
  {https://ui.adsabs.harvard.edu/abs/2018MNRAS.475..648P} {475, 648}

\bibitem[\protect\citeauthoryear{Planck~Collaboration Ade
  et~al.,}{Planck~Collaboration et~al.}{2016}]{Planck2016}
Planck~Collaboration Ade P.~A.,  et~al., 2016, Astronomy \& Astrophysics, 594,
  A13

\bibitem[\protect\citeauthoryear{{Ploeckinger}, {Sharma}, {Schaye}, {Crain},
  {Schaller}  \& {Barber}}{{Ploeckinger} et~al.}{2018}]{Ploeckinger2018}
{Ploeckinger} S.,  {Sharma} K.,  {Schaye} J.,  {Crain} R.~A.,  {Schaller} M.,
  {Barber} C.,  2018, \mn@doi [\mnras] {10.1093/mnras/stx2787}, \href
  {https://ui.adsabs.harvard.edu/abs/2018MNRAS.474..580P} {474, 580}

\bibitem[\protect\citeauthoryear{{Read}, {Wilkinson}, {Evans}, {Gilmore}  \&
  {Kleyna}}{{Read} et~al.}{2006}]{Read2006}
{Read} J.~I.,  {Wilkinson} M.~I.,  {Evans} N.~W.,  {Gilmore} G.,   {Kleyna}
  J.~T.,  2006, \mn@doi [\mnras] {10.1111/j.1365-2966.2005.09861.x}, \href
  {https://ui.adsabs.harvard.edu/abs/2006MNRAS.366..429R} {366, 429}

\bibitem[\protect\citeauthoryear{{Rodriguez} \& {Merch{\'a}n}}{{Rodriguez} \&
  {Merch{\'a}n}}{2020}]{Rodriguez2020}
{Rodriguez} F.,  {Merch{\'a}n} M.,  2020, \mn@doi [\aap]
  {10.1051/0004-6361/201937423}, \href
  {https://ui.adsabs.harvard.edu/abs/2020A&A...636A..61R} {636, A61}

\bibitem[\protect\citeauthoryear{{Rodriguez-Gomez} et~al.,}{{Rodriguez-Gomez}
  et~al.}{2015}]{Rodriguez-Gomez2015}
{Rodriguez-Gomez} V.,  et~al., 2015, \mn@doi [\mnras] {10.1093/mnras/stv264},
  \href {https://ui.adsabs.harvard.edu/abs/2015MNRAS.449...49R} {449, 49}

\bibitem[\protect\citeauthoryear{{Rodriguez}, {Montero-Dorta}, {Angulo},
  {Artale}  \& {Merch{\'a}n}}{{Rodriguez} et~al.}{2021}]{Rodriguez2021}
{Rodriguez} F.,  {Montero-Dorta} A.~D.,  {Angulo} R.~E.,  {Artale} M.~C.,
  {Merch{\'a}n} M.,  2021, \mn@doi [\mnras] {10.1093/mnras/stab1571}, \href
  {https://ui.adsabs.harvard.edu/abs/2021MNRAS.505.3192R} {505, 3192}

\bibitem[\protect\citeauthoryear{Rodr{\'\i}guez, Garcia~Lambas, Padilla,
  Tissera, Bignone, Dominguez-Tenreiro, Gonzalez  \& Pedrosa}{Rodr{\'\i}guez
  et~al.}{2022}]{RodriguezSilvio2022}
Rodr{\'\i}guez S.,  Garcia~Lambas D.,  Padilla N.,  Tissera P.,  Bignone L.,
  Dominguez-Tenreiro R.,  Gonzalez R.,   Pedrosa S.,  2022, Monthly Notices of
  the Royal Astronomical Society, 514, 6157

\bibitem[\protect\citeauthoryear{{Safarzadeh} \& {Loeb}}{{Safarzadeh} \&
  {Loeb}}{2019}]{Safarzadeh2019}
{Safarzadeh} M.,  {Loeb} A.,  2019, \mn@doi [\mnras] {10.1093/mnrasl/slz053},
  \href {https://ui.adsabs.harvard.edu/abs/2019MNRAS.486L..26S} {486, L26}

\bibitem[\protect\citeauthoryear{{Sales}, {Navarro}, {Pe{\~n}afiel}, {Peng},
  {Lim}  \& {Hernquist}}{{Sales} et~al.}{2020}]{Sales2020}
{Sales} L.~V.,  {Navarro} J.~F.,  {Pe{\~n}afiel} L.,  {Peng} E.~W.,  {Lim} S.,
   {Hernquist} L.,  2020, \mn@doi [\mnras] {10.1093/mnras/staa854}, \href
  {https://ui.adsabs.harvard.edu/abs/2020MNRAS.494.1848S} {494, 1848}

\bibitem[\protect\citeauthoryear{{Shi} et~al.,}{{Shi} et~al.}{2020}]{Shi2020}
{Shi} J.,  et~al., 2020, \mn@doi [\apj] {10.3847/1538-4357/ab8464}, \href
  {https://ui.adsabs.harvard.edu/abs/2020ApJ...893..139S} {893, 139}

\bibitem[\protect\citeauthoryear{{Simpson}, {Grand}, {G{\'o}mez}, {Marinacci},
  {Pakmor}, {Springel}, {Campbell}  \& {Frenk}}{{Simpson}
  et~al.}{2018}]{Simpson2018}
{Simpson} C.~M.,  {Grand} R. J.~J.,  {G{\'o}mez} F.~A.,  {Marinacci} F.,
  {Pakmor} R.,  {Springel} V.,  {Campbell} D. J.~R.,   {Frenk} C.~S.,  2018,
  \mn@doi [\mnras] {10.1093/mnras/sty774}, \href
  {https://ui.adsabs.harvard.edu/abs/2018MNRAS.478..548S} {478, 548}

\bibitem[\protect\citeauthoryear{{Smith}, {Davies}  \& {Nelson}}{{Smith}
  et~al.}{2010}]{Smith2010}
{Smith} R.,  {Davies} J.~I.,   {Nelson} A.~H.,  2010, \mn@doi [\mnras]
  {10.1111/j.1365-2966.2010.16545.x}, \href
  {https://ui.adsabs.harvard.edu/abs/2010MNRAS.405.1723S} {405, 1723}

\bibitem[\protect\citeauthoryear{{Smith}, {Choi}, {Lee}, {Rhee},
  {Sanchez-Janssen}  \& {Yi}}{{Smith} et~al.}{2016}]{Smith2016}
{Smith} R.,  {Choi} H.,  {Lee} J.,  {Rhee} J.,  {Sanchez-Janssen} R.,   {Yi}
  S.~K.,  2016, \mn@doi [\apj] {10.3847/1538-4357/833/1/109}, \href
  {https://ui.adsabs.harvard.edu/abs/2016ApJ...833..109S} {833, 109}

\bibitem[\protect\citeauthoryear{Somerville \& Dav{\'e}}{Somerville \&
  Dav{\'e}}{2015}]{Somerville2015}
Somerville R.~S.,  Dav{\'e} R.,  2015, Annual Review of Astronomy and
  Astrophysics, 53, 51

\bibitem[\protect\citeauthoryear{{Springel}}{{Springel}}{2010}]{Springel2010}
{Springel} V.,  2010, \mn@doi [\mnras] {10.1111/j.1365-2966.2009.15715.x},
  \href {https://ui.adsabs.harvard.edu/abs/2010MNRAS.401..791S} {401, 791}

\bibitem[\protect\citeauthoryear{{Springel}, {White}, {Tormen}  \&
  {Kauffmann}}{{Springel} et~al.}{2001}]{Springel2001}
{Springel} V.,  {White} S. D.~M.,  {Tormen} G.,   {Kauffmann} G.,  2001,
  \mn@doi [\mnras] {10.1046/j.1365-8711.2001.04912.x}, \href
  {https://ui.adsabs.harvard.edu/abs/2001MNRAS.328..726S} {328, 726}

\bibitem[\protect\citeauthoryear{{Springel} et~al.,}{{Springel}
  et~al.}{2018}]{Springel2018}
{Springel} V.,  et~al., 2018, \mn@doi [\mnras] {10.1093/mnras/stx3304}, \href
  {https://ui.adsabs.harvard.edu/abs/2018MNRAS.475..676S} {475, 676}

\bibitem[\protect\citeauthoryear{{Springel}, {Pakmor}, {Zier}  \&
  {Reinecke}}{{Springel} et~al.}{2021}]{Springel2021}
{Springel} V.,  {Pakmor} R.,  {Zier} O.,   {Reinecke} M.,  2021, \mn@doi
  [\mnras] {10.1093/mnras/stab1855}, \href
  {https://ui.adsabs.harvard.edu/abs/2021MNRAS.506.2871S} {506, 2871}

\bibitem[\protect\citeauthoryear{Valluri}{Valluri}{1993}]{Valluri1993}
Valluri M.,  1993, The Astrophysical Journal, 408, 57

\bibitem[\protect\citeauthoryear{{Vogelsberger} et~al.,}{{Vogelsberger}
  et~al.}{2018}]{Vogelsberger2018}
{Vogelsberger} M.,  et~al., 2018, \mn@doi [\mnras] {10.1093/mnras/stx2955},
  \href {https://ui.adsabs.harvard.edu/abs/2018MNRAS.474.2073V} {474, 2073}

\bibitem[\protect\citeauthoryear{{Vogelsberger} et~al.,}{{Vogelsberger}
  et~al.}{2020}]{Vogelsberger2020}
{Vogelsberger} M.,  et~al., 2020, \mn@doi [\mnras] {10.1093/mnras/staa137},
  \href {https://ui.adsabs.harvard.edu/abs/2020MNRAS.492.5167V} {492, 5167}

\bibitem[\protect\citeauthoryear{Vollmer, Cayatte, Balkowski  \&
  Duschl}{Vollmer et~al.}{2001}]{Vollmer2001}
Vollmer B.,  Cayatte V.,  Balkowski C.,   Duschl W.,  2001, The Astrophysical
  Journal, 561, 708

\bibitem[\protect\citeauthoryear{{Vulcani} et~al.,}{{Vulcani}
  et~al.}{2018}]{Vulcani2018}
{Vulcani} B.,  et~al., 2018, \mn@doi [\apjl] {10.3847/2041-8213/aae68b}, \href
  {https://ui.adsabs.harvard.edu/abs/2018ApJ...866L..25V} {866, L25}

\bibitem[\protect\citeauthoryear{{White} \& {Frenk}}{{White} \&
  {Frenk}}{1991}]{WhiteFrenk1991}
{White} S. D.~M.,  {Frenk} C.~S.,  1991, \mn@doi [\apj] {10.1086/170483}, \href
  {https://ui.adsabs.harvard.edu/abs/1991ApJ...379...52W} {379, 52}

\bibitem[\protect\citeauthoryear{{White} \& {Rees}}{{White} \&
  {Rees}}{1978}]{WhiteRees1978}
{White} S.~D.~M.,  {Rees} M.~J.,  1978, \mn@doi [\mnras]
  {10.1093/mnras/183.3.341}, \href
  {https://ui.adsabs.harvard.edu/abs/1978MNRAS.183..341W} {183, 341}

\bibitem[\protect\citeauthoryear{Yang, Mo  \& van~den Bosch}{Yang
  et~al.}{2009}]{Yang2009}
Yang X.,  Mo H.~J.,   van~den Bosch F.~C.,  2009, \mn@doi [The Astrophysical
  Journal] {10.1088/0004-637x/693/1/830}, 693, 830

\bibitem[\protect\citeauthoryear{Zhang, Yang  \& Guo}{Zhang
  et~al.}{2021}]{Zhang2021}
Zhang Y.,  Yang X.,   Guo H.,  2021, Monthly Notices of the Royal Astronomical
  Society, 507, 5320

\bibitem[\protect\citeauthoryear{{van Dokkum} et~al.,}{{van Dokkum}
  et~al.}{2018}]{vD2018_noDM1}
{van Dokkum} P.,  et~al., 2018, \mn@doi [\nat] {10.1038/nature25767}, \href
  {https://ui.adsabs.harvard.edu/abs/2018Natur.555..629V} {555, 629}

\bibitem[\protect\citeauthoryear{{van Dokkum}, {Danieli}, {Abraham}, {Conroy}
  \& {Romanowsky}}{{van Dokkum} et~al.}{2019}]{vD2019_noDM2}
{van Dokkum} P.,  {Danieli} S.,  {Abraham} R.,  {Conroy} C.,   {Romanowsky}
  A.~J.,  2019, \mn@doi [\apjl] {10.3847/2041-8213/ab0d92}, \href
  {https://ui.adsabs.harvard.edu/abs/2019ApJ...874L...5V} {874, L5}

\bibitem[\protect\citeauthoryear{{van Dokkum} et~al.,}{{van Dokkum}
  et~al.}{2022}]{vD2022}
{van Dokkum} P.,  et~al., 2022, \mn@doi [\nat] {10.1038/s41586-022-04665-6},
  \href {https://ui.adsabs.harvard.edu/abs/2022Natur.605..435V} {605, 435}

\makeatother
\end{thebibliography}





\bsp	
\label{lastpage}
\end{document}